\documentclass[aps,prb,showpacs,twocolumn,amsmath,amssymb,superscriptaddress]{revtex4}

\usepackage{graphicx}
\usepackage{multirow}
\usepackage{dcolumn}
\usepackage{amssymb,amscd,xypic,bm,dsfont,wasysym}
\usepackage{float}
\usepackage{color}
\usepackage{xcolor}

\newcommand{\footstar}[1]{$^*$ \footnotetext{$^*$#1}}

\begin{document}

\title{Topological Superconductivity induced by Ferromagnetic Metal Chains}

\author{Jian Li$^*$} 
   \affiliation{Department of Physics, Princeton University, Princeton, NJ 08544, USA}
\author{Hua Chen\footstar{These authors contributed equally to this work}}
    \affiliation{Department of Physics, University of Texas at Austin, Austin, TX 78712, USA}
\author{Ilya K. Drozdov} 
   \affiliation{Department of Physics, Princeton University, Princeton, NJ 08544, USA}
\author{A. Yazdani}
    \affiliation{Department of Physics, Princeton University, Princeton, NJ 08544, USA}
\author{B. Andrei Bernevig}
    \affiliation{Department of Physics, Princeton University, Princeton, NJ 08544, USA}
\author{A.H. MacDonald}
    \affiliation{Department of Physics, University of Texas at Austin, Austin, TX 78712, USA}
    
\begin{abstract}
Recent experiments have provided evidence that one-dimensional (1D) topological superconductivity
can be realized experimentally by placing transition metal atoms that form a ferromagnetic chain
on a superconducting substrate.  We address some properties of this type of systems by using
Slater-Koster tight-binding model to account for important features of the electronic structure
of the transition transition metal chains on the superconducting substrate.
We predict that topological superconductivity is nearly universal when ferromagnetic transition metal chains 
form straight lines on superconducting substrates and
that it is possible for more complex chain structures.  When the chain is weakly coupled to the substrate 
and is longer than superconducting coherence lengths,
its proximity induced superconducting gap is $\sim \Delta E_{so} / J$ where $\Delta$ is the $s$-wave pair-potential on the 
chain, $E_{so}$ is the spin-orbit splitting energy induced in the normal chain state bands by hybridization with the superconducting substrate, 
and $J$ is the exchange-splitting of the ferromagnetic chain $d$-bands.  
Because of the topological character of the 1D superconducting state, 
Majorana end modes appear within the gaps of finite length chains.  We find, in agreement with the experiment, that when 
the chain and substrate orbitals are strongly hybridized, Majorana end
modes are substantially reduced in amplitude when separated from the chain end 
by less than the coherence length defined by the $p$-wave superconducting gap.
We conclude that Pb is a particularly favorable substrate material for ferromagnetic chain topological superconductivity because 
it provides both strong $s-$wave pairing and strong Rashba spin-orbit coupling, but that there is  
an opportunity to optimize properties by varying the atomic composition and structure of the chain.  
Finally, we note that in the absence of disorder a new chain magnetic symmetry, 
one that is also present in the crystalline topological insulators,
can stabilize multiple Majorana modes at the end of a single chain. 
\end{abstract}


\maketitle

\section{Introduction}
Recent interest in exploiting the exchange properties of Majorana states\cite{majorana_1937}
in $p$-wave superconductors\cite{moore_1991,kitaev_2001,ivanov_2001} 
as a basis for more robust quantum computation\cite{nayak_2008,alicea_2012} has motivated the invention 
of a variety of different strategies which can in principle be used to engineer topological superconductivity.\cite{fu_2008} 
One-dimensional topological superconductivity can be achieved by combining spin-orbit coupling with 
broken time reversal symmetry in a variety of different ways
to create effective $p$-wave superconductors.
Ideas have been proposed based on quantum spin Hall edge states, \cite{fu_2009}
semiconductor quantum wires, \cite{lutchyn_2010,oreg_2010}
half-metallic ferromagnets, \cite{Duckheim_2011, Chung_2011}
topological insulator nanowires, \cite{Cook_2011}
metallic chains, \cite{Potter_2012} 
strongly spin-orbit coupled superconductors, and 
helical magnetic chains. \cite{Choy11, ivar_2012, stevan_2013,pientka_2013,klinovaja_2013,braunecker_2013,vazifeh_2013}
Indeed there is strong, but at present still inconclusive, evidence\cite{mourik_2012,deng_2012,das_2012} that Majorana states 
have been realized by following the semiconductor nanowire strategy.  
The present work is motivated by the appearance\cite{MajoranaCenter,NobelSymposium} 
of telltale zero-bias anomalies in experimental work that was originally motivated by the helical magnetic chain idea,
but finally interpreted\cite{NobelSymposium,Nadj-Perge14} in terms of the properties of ferromagnetic chains.  
Ref.~\onlinecite{Nadj-Perge14} reports strong evidence that Fe chains on Pb are ferromagnetic,  that they  
are one-dimensional topological superconductors, and that Majorana end states are responsible for 
zero-bias anomalies in the local density of states measured near the ends of finite length chains. 
In this article we explain why topological states are not only possible, but for some structures overwhelmingly likely, 
when atomic chains formed
from late $3d$ transition elements (or other strong magnetic materials) are placed on the surface of a superconductor.  

The ubiquity of topological states is related to features in the electronic structure of straight transition metal chains detailed later. 
In order to bring out the essential physics in a transparent fashion we first study a simplified but still realistic chain model with 
proximity-induced $s$-wave pairing and $d$-orbital Slater-Koster tight-binding bands. We then model the case of a 
one-dimensional ferromagnetic chain embedded in the $(110)$ surface of bulk Pb,
the situation studied experimentally in Ref.~\onlinecite{Nadj-Perge14}.
We provide quantitative results for substrate induced spin-orbit coupling 
on the chain, for the superconducting gap of the chain, for the structure of Shiba states in this system, 
and for the spatial decay properties of Majorana states localized at the chain ends. 
Importantly we find that iron chains on Pb substrates are partially submerged beneath 
the surface, that the chain and substrate orbitals are strongly hybridized, and
that spatial decay of Majorana end modes along the chain can 
consequently occur on length scales shorter than the coherence length associated 
with the $p$-wave superconducting gap induced in the 
chain.  Finally, we also point out that in the absence of disorder, a combined magnetic symmetry (mirror times time-reversal) first identified in Ref.~\onlinecite{fang_2014} can stabilize multiple Majoranas at the end of a single Fe chain. 
\begin{figure}[h]
 \begin{center}
\includegraphics[width=3.6 in]{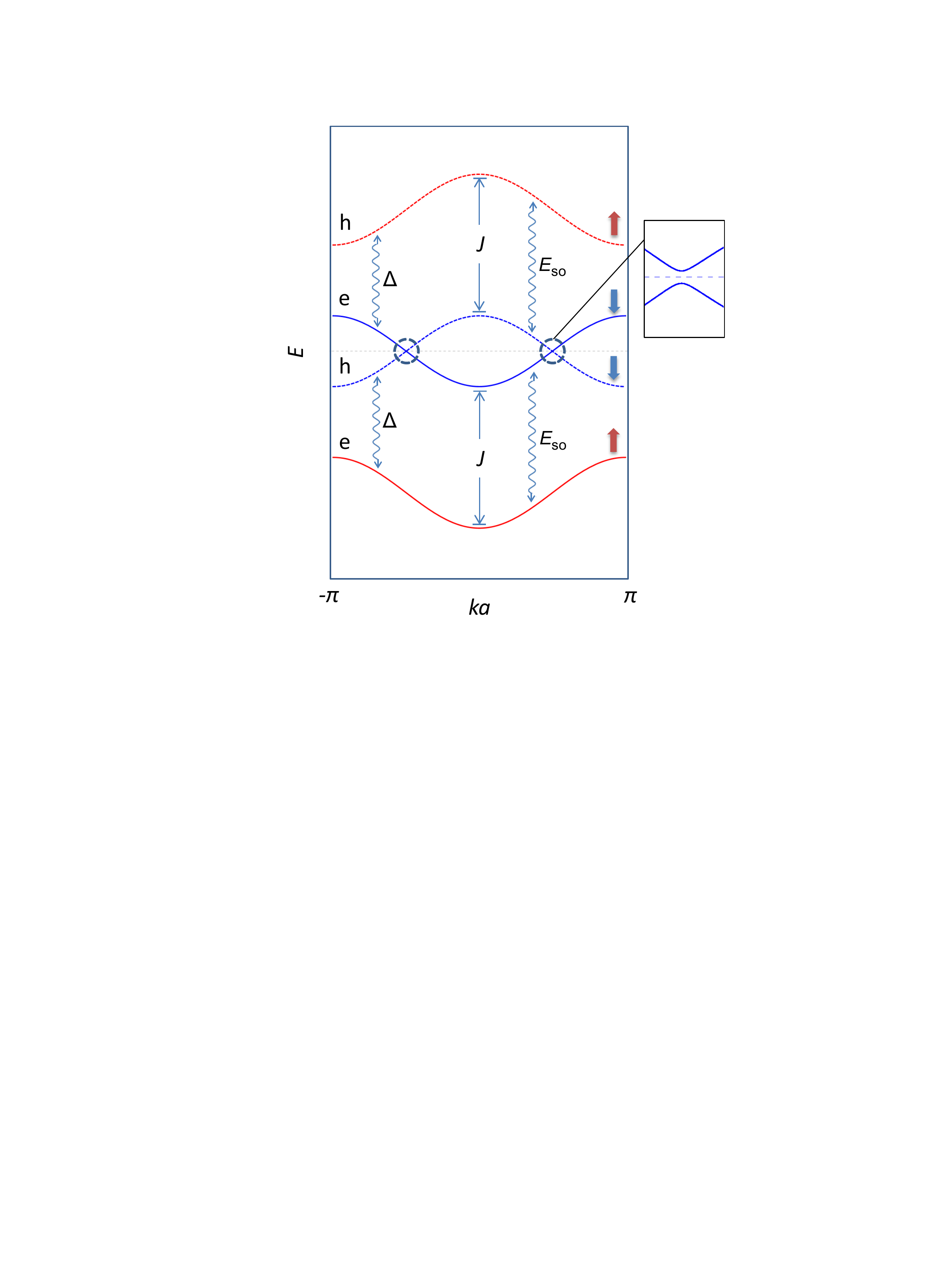}
 \end{center}
 \caption{Bogoliubov quasiparticle bands of a system with strong exchange splitting strength $J$,
and pair potential $\Delta$ and spin-orbit coupling strengths $E_{so}$ that are by comparison weaker.  
This illustration assumes that the majority-spin $d$-bands (red) are full and the 
minority-spin $d$-bands (blue) are partially filled, the usual case for transition metal ferromagnets.
The minority-spin electron (solid) and hole (dashed) bands which cross at the Fermi level are coupled via 
a virtual process in which the pair potential $\Delta$ couples minority spin electrons (holes) to
majority spin holes (electrons) and $E_{so}$ couples minority spin holes (electrons) to 
majority spin holes (electrons).  It follows that the quasiparticle gap at the Fermi energy
indicated in the inset is $\sim \Delta E_{so}/J$.}
\label{fig:schematic}
\end{figure}

Under most circumstances ferromagnetism and superconductivity are antagonistic.\cite{ferrosuperreview}  Superconductivity is however 
able to survive on a ferromagnetic chain because a single row of aligned spins does not generate significant magnetic 
flux density, obviating damaging orbital effects, and because the substrate provides a non-magnetic Cooper pair reservoir. 
The mean-field Hamiltonian of the ferromagnetic chain contains two spin-dependent terms, a very 
large spin-splitting term produced by magnetic order which is odd under time reversal, and a 
much smaller spin-orbit coupling term that is even under time reversal.  When only the large term is 
retained, quasiparticle wave functions are unperturbed and spin $\uparrow$ and $\downarrow$ quasiparticle
energies are shifted in opposite directions by half the exchange splitting $J$.  When the 
Fermi level lies in the minority spin bands (see Fig.~\ref{fig:schematic}), the electron and hole Bogoliubov bands 
which cross at the Fermi level have the same spin, the pair potential $\Delta$ couples quasiparticles with 
different bare energies, and the pair amplitude on the chain is small. The reservoir of Cooper pairs in the substrate 
effectively allows superconductivity to survive in the chain when it would be suppressed in a bulk system. 
Spin-orbit interactions produce a gap at the Fermi level because like-spin electrons and holes are coupled 
by a virtual process in which the pair potential reverses both spin and electron/hole character,
whereas spin-orbit coupling reverses spin without reversing charge. It follows that the gap at the Fermi energy is $\sim \Delta E_{so}/J$
where $E_{so}$ is the spin-orbit coupling strength.\cite{alicea_2011} The effective spin-orbit interaction matrix elements responsible for the gap 
are closely related to the pair creation and annihilation terms which were already carefully analyzed in the original BCS paper.\cite{bcs_1957,tinkham_1996} Pb substrates are rather unique in providing both relatively strong $s$-wave pairing and 
strong spin-orbit coupling. Because $E_{so}$ is not as small compared to $J$ as $\Delta$, at least in
systems with a Pb substrate, sizable Fermi level gaps are possible even though the Clogston\cite{clogston_1962} limit is
enormously exceeded, {\it i.e.} $J \gg \Delta$. The main focus of this paper is on explaining why this gapped superconducting state 
is topological more often than not. 
The system-parameter regime over which topological superconductivity can appear
is wider than that for most previously studied mechanisms for effective $p$-wave 
superconductivity. The material in this paper expands on theoretical ideas that were partially
presented in Ref.\onlinecite{Nadj-Perge14}.
 
The paper is organized as follows: In Section II we address the electronic structure of isolated transition metal chains and discuss how it is altered by proximity induced superconductivity. We explain why straight ferromagnetic transition metal chains almost always exhibit topological superconductivity and show why Rashba spin-orbit interactions, allowed in systems with broken inversion symmetry, are necessary to open a superconducting gap in the system.  In Section III we look at more realistic chain geometric configurations similar to the ones appearing in the experiment 
described in Ref.\onlinecite{Nadj-Perge14} and calculate their phase diagram when they are suspended 
and influenced by a singlet pair-potential whose strength is treated as a phenomenological parameter. 
In Section IV we consider the experimental situation of one dimensional Fe chains on the surface of a Pb superconductor to
which it is strongly hybridized.  We model the Pb substrate 
using a realistic tight-binding Hamiltonian with parameter values obtained from {\it ab-initio} calculations. 
We also identify a new magnetic symmetry that can protect more than one Majorana at one end of the chain, and construct a phase diagram 
for the number of Majorana modes per end. 
We also calculate the spatial extent of the Majorana states and show that the in the strongly
hybridized case the Majorana state amplitude exhibits 
strong deviations from the simple exponential decay of a suspended one-dimensional chain.
Finally, in Section V we present our conclusions. 

\section{Superconductivity in Ferromagnetic Chains}

\subsection{Slater-Koster model of a superconducting ferromagnetic chain}

Our discussion of topological superconductivity in ferromagnetic metal chains is 
informed by realistic electronic structure considerations. Since metallic ferromagnetism
is most often associated with Fermi level $d$-electrons, we focus our attention here 
on chains formed by transition metal atoms. Chains formed by rare earth atoms like Gd could 
however also be of interest. We first discuss the properties of band Hamiltonians $H_{0}$ 
with $d$-orbital Slater-Koster approximation tight-binding ($H_{SK}$),
Stoner-theory spin-splitting ($H_{J}$), and atomic-like
spin-orbit coupling ($H_{so}$) contributions:
\begin{eqnarray}\label{eq:Hchain}
H_{0}=H_{SK}+H_{J}+H_{so}.  
\end{eqnarray}
Of the three terms in the band Hamiltonian, only the hopping term $H_{SK}$ is spin-independent:
\begin{eqnarray}
H_{SK}=\sum_{\langle ij\rangle \alpha' \alpha \sigma} \; t_{\alpha'\alpha} \; c_{i\alpha'\sigma}^\dag c_{j\alpha\sigma}.
\end{eqnarray}
Here $i$ and $j$ label sites, $\langle ij\rangle$ implies a restriction to nearest neighbor sites, 
$\sigma$ labels spin, and $\alpha'\alpha$ label the five $d$-orbitals on each site.
As will become clear later neither the inclusion of $s$-orbitals, which are not strongly spin-polarized according to 
{\it ab initio} calculations\cite{Nadj-Perge14}, nor the inclusion of longer-range hopping processes would modify our main 
conclusions. The $t_{\alpha '\alpha}$ hopping parameters are real Slater-Koster integrals that 
depend for each orbital pair on the direction cosines of the vector
connecting nearest neighbors, and on the three Slater-Koster parameters $V_{dd\sigma}$, $V_{dd\pi}$, and $V_{dd\delta}$.  
We focus our attention first on straight chains, using this geometry to identify important trends.
Real chains need not be straight \cite{Nadj-Perge14} or,   because of incommensurability
between the isolated chain and substrate lattice constants, even periodic.  However we expect that straight chain features 
in the electronic structure will sometimes be reflected in actual geometries.  For concrete calculations we use the Slater-Koster parameter values 
proposed for bulk Fe in Ref.\onlinecite{zhong_1993}, which are listed in Table~\ref{tab:Feparameters} for completeness.
These parameters exhibit the generic\cite{andersen_1975} metallic band property $|V_{dd\sigma}| > |V_{dd\pi}| > |V_{dd\delta}|$ 
which we will see is key to the ubiquity of topological states in straight chains.  

\begin{table}[h]
\caption{Slater-Koster tight-binding model parameters for Fe (in eV). 
The hopping integral values are for the nearest-neighbor distance of bulk Fe(bcc), $r_0=2.383$ \AA.}
\centering
$\,$\\
\begin{tabular}{cc}
\hline\hline
Parameters & Value (eV) \\\hline
$V_{dd\sigma}$ & -0.6702 \\
$V_{dd\pi}$ & 0.5760 \\
$V_{dd\delta}$ & -0.1445 \\
\hline\hline
 \end{tabular}\label{tab:Feparameters}
\end{table}

We first consider models in which both spin-dependent terms  
$H_{J}$ and $H_{so}$ are diagonal in site: 
\begin{eqnarray}
&&H_{J}=- J \, \hat{m} \cdot {\bm s},\\\nonumber
&&H_{so}=\lambda_{so} \, \bm L \cdot {\bm s},
\end{eqnarray}
where $J$ is the ferromagnetic state quasiparticle spin-splitting energy, $\hat{m}$ is the magnetization direction
on the chain, $\lambda_{so}$ is the spin-orbit coupling parameter, and $\bm L$ and $\bm s$ are respectively the atomic angular momentum 
and electron spin operators. It will be important in what follows that $H_{J}$ changes sign under time reversal 
whereas $H_{so}$ is time-reversal invariant. For Fe $J \sim 2.5$ eV and $\lambda_{so} \sim 0.06$ eV.
By comparing with {\it ab initio} electronic structure calculations one can confirm that this simple model
accounts accurately for the electronic structure and magnetic anisotropy of isolated Fe chains. 
We will see later that in straight chains  the $E_{so}$ coupling required in
Fig.~\ref{fig:schematic} to produce gaps is not provided by atomic spin-orbit coupling.  
This observation elevates the importance of spin-orbit coupling inherited from the superconducting 
substrate through orbital hybridization.   

\begin{figure}[h]
 \begin{center}
\includegraphics[width=2.5 in]{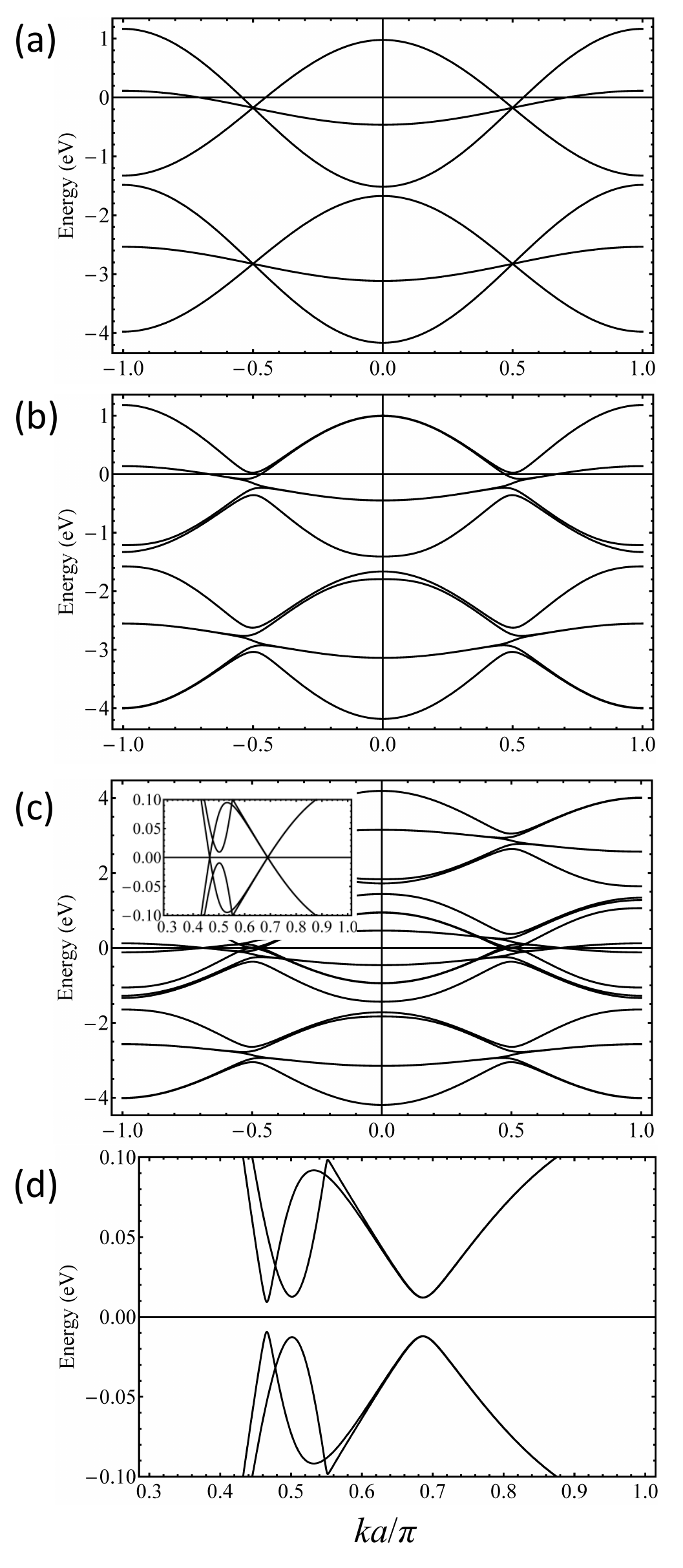}
\end{center}
\caption{Model band structures for straight Fe chain. 
(a) $\lambda_{SO}=0$.  (b) $\lambda_{SO}=0.2$ eV. (c) BdG spectrum with 
$\lambda_{SO}=0.2$ eV and $\Delta = 0.2$ eV.  The pair-potential 
value used in this illustration is unrealistically large and has 
been chosen for easy visualization.  
The inset highlights the quasiparticle bands which cross
at the Fermi energy. (d) Same as (c) but with an 
orbital independent Rashba spin-orbit term with coupling constant $t_R=0.05$ eV. (See text)}
\label{fig:2}
\end{figure}

Figure~\ref{fig:2} (a) shows the band structure of a straight Fe chain without atomic spin-orbit coupling. 
Due to rotational symmetry around the chain direction ($\hat{x}$), there are two pairs of spin-degenerate 
bands, two $dd\pi$ bands ($zx$ and $xy$ orbitals) with minima at $ka=\pi$ 
and two narrower $dd\delta$ bands ($yz$ and $y^2-z^2$ orbitals) with minima at $ka=0$.  
The broadest $dd\sigma$ band is not orbitally degenerate and also has its minimum at the zone center.
Because the spin-splitting exceeds the chain band width, which is smaller than the bulk band width because 
of the reduced coordination number in a 1D system, the minority and majority spin $d$-bands do not overlap.  
When spin-orbit coupling ($H_{so}$) is added (Figure~\ref{fig:2} (b)),
with $\hat{m} \cdot \hat{x} = 0$, corresponding to an easy magnetization direction perpendicular 
to the chain~\cite{footnote_anisotropy,Nadj-Perge14}, states near the Fermi level of a late transition metal system are still nearly pure 
minority spin in character and the two-fold degeneracy of the $dd\delta$ and $dd\pi$ bands 
is only weakly lifted. Neglecting this small splitting, the number of minority spin bands which cross 
the Fermi level is always odd. This property will be responsible for superconductivity that 
is always topological, provided that spin-orbit coupling mixes the superconducting quasiparticle states  
which cross at the exchange-shifted Fermi energy. Straight transition metal chains are 
therefore favorable for topological superconductivity.  (Since the Fe chains in the initial 
experimental studies~\cite{Nadj-Perge14} were not straight, this observation suggests one strategy to follow in an effort to make progress in improving the magnetic-chain Majorana platform.)  In Sec.~\ref{sec:MS_chain} we will provide a more detailed analysis of non-straight (zigzag) chains in terms of their topological properties. 

We now add a pair-potential term to the Hamiltonian, assuming that 
it is dominated by a local, orbital-independent, spin-singlet contribution:
\begin{eqnarray}\label{eq:hpair}
H_{ pair}=\Delta\sum_{\alpha }\left (c_{\alpha\uparrow}^\dag c_{\alpha\downarrow}^\dag + c_{\alpha\downarrow} c_{\alpha\uparrow} \right ),
\end{eqnarray}
where we have chosen a real Slater-Koster basis for the $d$-orbitals. 
Although outside the scope of the present work, it will also be interesting to consider 
spin-triplet contributions to the pair potential, which is inevitably present in the presence of broken inversion symmetry and strong spin-orbit coupling.
We neglect it in the same spirit as we neglect longer range hopping on the chain, {\it i.e.} as an expedient to reduce the number of parameters in our model calculations. We have not identified a 
mechanism by which weak triplet pairing would alter our main conclusions. Fully realistic calculations 
of pair-potentials would have to account for modifications of phonons and electron-phonon coupling near the surface of the superconducting substrate. Although these calculations are feasible, we judge that it would be premature to undertake this effort at present.  

In its doubled particle-hole Nambu space the $20 N \times 20 N$ Bogoliubov-de Gennes (BdG) mean-field 
Hamiltonian\cite{ketterson_1999} for a chain with singlet-pairing and $N$ sites is 
\begin{eqnarray}\label{eq:BdG}
{\cal H}_{BdG} = \left( \begin{array}{cc}
H_{chain} & \Delta \;  \mathbf{I}_{5N\times 5N} \otimes \, i \sigma_y  \\
-\Delta \; \mathbf{I}_{5N\times 5N} \otimes \, i \sigma_y & -H^*_{chain}
\end{array} \right)
\end{eqnarray}
where $\sigma_y$ is a Pauli matrix acting on spin labels.

The spectrum obtained by diagonalizing ${\cal H}_{BdG}$ 
is illustrated in Figs.~\ref{fig:2} (c).  
Interestingly no gap is opened at the Fermi level, indicating that the singlet-pairing induced virtual
coupling between minority spin electrons and holes vanishes.
This property can be traced to the charge conjugation symmetry of the 
BdG equations combined with the inversion symmetry of the model described thus far, as explained in detail in Subsection ~\ref{gapless} below.  
Spin-orbit coupling within the chain, which usually provides the largest spin-orbit coupling scale, does not support the formation of a gapped topological state unless the chain structure breaks inversion symmetry.
We conclude that chain structures that break inversion symmetry can potentially be favorable for topological state formation. Fortunately, inversion symmetry is always broken for chains which lie on the surface of a substrate. At a surface hopping processes in which 
the spin-component perpendicular to the surface is flipped are always allowed to depend on 
hopping direction, leading to band Hamiltonian terms that are odd in momentum.
This effect is generically referred to as Rashba spin-orbit coupling.  
We therefore add a band-independent term of the form,
\begin{eqnarray}\label{eq:rashbasimple}
H_{R}=it_R\sum_{\langle ij\rangle \gamma\tau} c_{i\gamma}^\dag c_{j\tau} (\hat{d}_{ij}\times \bm \sigma_{\gamma\tau})\cdot\hat{z},
\end{eqnarray}
to the toy model Hamiltonian. 
Here $\hat{d}_{ij}$ is a unit vector pointing from site $i$ to site $j$, and $\gamma$, $\tau$ are spin indices. 
As shown in Fig.~\ref{fig:2} (d), as soon as $t_R$ becomes nonzero a gap opens at the Fermi level. 
We conclude that chains on the surface of a superconductor should generally have more robust topological states than submerged chains, because they have stronger inversion symmetry breaking and should therefore generally 
have stronger Rashba spin-orbit interactions. The Rashba process is discussed in more detail in Subsection ~\ref{Rashba} below.  

\subsection{Inversion-Symmetry and Finite Gaps}
\label{gapless}

In this subsection we explain the observation made in the previous subsection that inversion symmetry protects gapless points in 
one dimensional spinful charge-conjugation-symmetric systems. 
We will show that when inversion and spinful charge conjugation symmetry are both present in 1D, gapless Fermi points are stable. Adding inversion symmetry to the BdG equation leads to an analog of Weyl fermions in 3D, which however do not need symmetry to be protected, and of the 2D fermions in graphene, which need combined inversion and time reversal symmetry to be protected (in the absence of spin-orbit coupling) from gapping. In all these cases, the issue of whether or not gapped points are allowed can be addressed by considering an effective Hamiltonian including only the bands involved at the gapless crossing point, counting the number of symmetry allowed parameters in this reduced Hamiltonian, and checking to see whether or not it is larger than the space dimension of the system. When the number of allowed Hamiltonian parameters is equal to (or smaller) than the space dimension, momentum tuning parameters can be adjusted to points (or surfaces) at which the reduced Hamiltonian vanishes.  In this case level crossings are generally allowed and do not require fine-tuning of the Hamiltonian. 

Since the BdG Hamiltonian of a ferromagnetic chain always breaks time reversal symmetry, bands are singly degenerate at
generic points in the 1D Brillouin zone. In order to analyze a gap-closing transition, we have to consider a 1D $k$-dependent reduced Hamiltonian describing two Bogoliubov bands that are about to touch at zero energy due to charge conjugation. To examine whether 
or not spin-orbit coupling is almost certain to open a gap we expand the $2 \times 2$ reduced BdG Hamiltonian in terms of Pauli matrices:  
\begin{equation}\label{eq:effectiveHk}
H(k)= \sum_{i=1,2,3} d_i(k)  \sigma_i.  
\end{equation}
Inversion $P$ and charge conjugation $C$ operations transform
the Hamiltonian as follows:  
\begin{equation}
P H(k) P^{-1} = H(-k);\;\;\; CH(k) C^{-1} = - H^*(-k).
\end{equation}
Hence the little group of the Hamiltonian at $k$ is
\begin{equation}
(PC) H(k) (PC)^{-1} = - H^*(k)
\end{equation}
For spinful fermions the matrix $C$ has the property $(CK)^2=1$ where $K$ is complex conjugation. For fermions in the presence of SU(2) symmetry (no spin-orbit coupling), a basis rotation can be made in spin space to 
make $(CK)^2=-1$ (effectively spinless), but this is not a physical situation. 

For the spinful fermion case we can choose from several representations of the inversion and charge conjugation operators on the two 
crossing bands described by the reduced Hamiltonian in Eq.~\ref{eq:effectiveHk}. 
The only restrictions is that these operators satisfy the squaring relations discussed above, and the commutation relation $[P,CK] = 0$.  
Suppose the inversion operator is the identity operator $I$. 
Then for the $C$ operator we can choose $C=\sigma_x$ or $\sigma_z$. 
In the first case imposing the little group symmetry requires that 
$d_x(k)= d_y(k) =0 ,\;\; \forall k$, while in the second case it requires $d_z(k)= d_y(k) =0 ,\;\; \forall k$.
The Hamiltonian therefore has codimension zero. For example, the Hamiltonian for the first case is $d_z(k) \sigma_z$. 
A gapless point at some point  $K_0$ has $d_z(k) \sim (k-K_0)$. Adding a small $d_z$ term can only move the 1D Dirac point,
and cannot produce a gap. One can pick other representations of inversion and convince oneself that the Hamiltonian still has codimension zero. 
For example $P=\sigma_z$, $C= I$ is just a shuffling of the representation above.

For completeness we also discuss the situation of effectively spinless fermions [$(CK)^2=-1$]. The representation of $C$ in this case is $i \sigma_y$. Taking $P=I$ we have $  \sigma_y H(k) \sigma_y = - H^*(k)$  which 
does not impose any constraints on the Hamiltonian. The codimension in this case is 2, and the system is almost certainly gapped. 

\subsection{Rashba Spin-Orbit Coupling}
\label{Rashba}  

As explained above, inversion-symmetry breaking Rashba spin-orbit coupling is crucial to realize topological superconductivity in ferromagnetic chains with 1D inversion symmetry. Rashba spin-orbit coupling is always present in the supported chain system
because inversion symmetry is inevitably broken by the position of the chain on top of a substrate. 
Previous proposals for Majorana end modes in 1D chains have mainly focused on inversion symmetry breaking within the chains \cite{lutchyn_2010,oreg_2010}.  
For a 3$d$ ferromagnetic chain on Pb however, hybridization with the 
strongly spin-orbit coupled states of the substrate likely\cite{Nadj-Perge14} plays the dominant role.

\begin{figure}[h]
 \begin{center}
\includegraphics[width=0.4\textwidth]{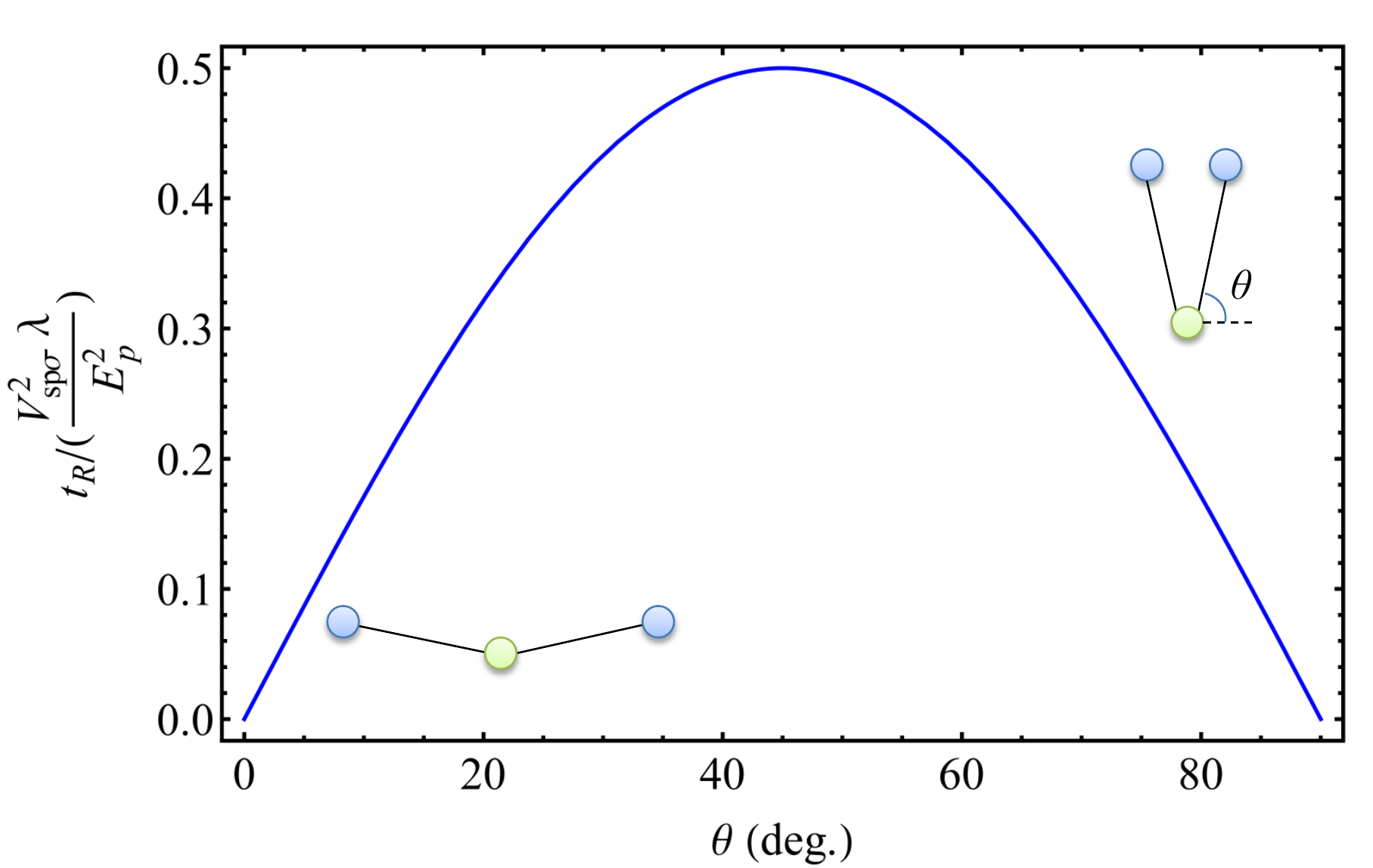}
 \end{center}
 \caption{Heuristic example of Rashba spin-orbit coupling induced by hopping on the chain via 
 substrate sites with strong spin-orbit coupling. In this illustration, the two-blue atoms can be associated with 
 neighboring atoms on the chain and the green atom with a Pb atom in the substrate. Inversion symmetry is broken
 because the substrate atom is below the chain atoms. The Rashba coupling 
 strength is proportional to the difference between left to right and right to left, spin-$\uparrow$ to spin$\downarrow$ virtual hopping between the chain atoms. The Rashba spin-orbit coupling for this simple toy model is plotted here as a function of the chain-substrate-chain bond angle.}
 \label{fig:4}
\end{figure}

In this section we discuss the physical processes leading to Rashba spin-orbit coupling in a ferromagnetic chain coupled to a substrate that has strong atomic spin-orbit coupling. We start from the heuristic example of two atoms with a single $s$-orbital, linked by an atom with only $p$-orbitals only, and assume that there is no direct hopping between the two $s$ atoms (Fig.~\ref{fig:4}). 
The choices of $s$- and $p$-orbitals are not essential and the argument below can be easily applied to other types of orbitals. The angle $\theta$ between the line defined by the two $s$-atoms and one $sp$ bond 
determines the extent of inversion symmetry breaking in this simple three-atom system. Choosing the zero of energy as the $s$-orbital site energy and assuming that the $s-p$ hybridization is weak, the Hamiltonian for virtual hopping between the two $s$ atoms from right to left via the $p$ atom is  
\begin{eqnarray}
T_{ss}=T^\dag_{sp}H_{p}^{-1}T_{sp},
\end{eqnarray} 
where $T_{sp}$ is the spin-independent but orbital-dependent hopping matrix between $s$ and $p$ orbitals proportional to the $V_{sp\sigma}$ Slater-Koster parameter, and $H_p$ is the local Hamiltonian of the $p$ atom 
including both a site energy $E_p$ and atomic spin-orbit coupling:
\begin{eqnarray}
H_{p}=E_{p}+\lambda \bm L \cdot \bm S.
\end{eqnarray}
$T_{ss}$ is a $2 \times 2$ matrix in which the first label is the $s$-orbital spin on the left site and the second label is the $s$-orbital spin on the right site. Because the overall Hamiltonian is Hermitian, the left to right hopping Hamiltonian can be obtained by reversing spin labels and 
taking a complex conjugate. Because the system lacks 3D inversion symmetry we expect a Rashba contribution to the effective hopping Hamiltonian, and characterize its strength by the coupling constant 
\begin{eqnarray}\label{eq:tRdef}
t_R= \frac{1}{2}(T_{s\uparrow,s\downarrow}-T_{s\downarrow,s\uparrow}^*).
\end{eqnarray}
Assuming $E_p$ to be much larger than $V_{sp\sigma}$ and $\lambda$, we arrive at the following expression for
the Rashba coupling constant of this illustrative toy model:
\begin{eqnarray}\label{eq:tRspmodel}
t_R=\frac{V_{sp\sigma}^2}{E_p} \frac{\lambda}{2 E_p}\sin(2\theta).
\end{eqnarray}

Although this simplified model does not apply directly to 
realistic transition metal chains on Pb substrates, 
there are several general remarks we can make based on Eq.~\ref{eq:tRspmodel}.
({\it i}): Rashba spin-orbit coupling is due to both atomic spin-orbit coupling and structural inversion symmetry breaking. 
Note, however, that even in this simple model $t_R$ is not a monotonic function of $\theta$. The Rashba spin-orbit coupling strength on a chain will always depend sensitively on the chain structure and on its coordination with the structure of the substrate. If these are known, it is a conceptually straightforward to calculate Rashba interactions quantitatively.    
({\it ii}):  $t_R$ should be roughly proportional to $t^2/ \delta E$, where $t$ is a typical hopping parameter between the system of interest
({\it e.g.}, an atomic chain) and the environment ({\it e.g.}, a substrate), and $\delta E$ is the energy difference between the system and the environment. 
This is easy to understand from a perturbation theory point of view. In general both $t$ and $\delta E$ can be matrices due to the presence of many orbitals. 
Specially, if the band structure of the system is diagonal in some localized Wannier orbital basis (such as in the straight Fe chain), different bands will in general acquire different Rashba spin-orbit coupling by interacting differently with the environment, in addition to possible orbital off-diagonal hopping. The largest Rashba spin-orbit coupling will be in the bands whose orbitals have strongest hybridization with the spin-orbit coupled environmental states. ({\it iii}):  The calculation described here includes only the lowest order process leading to Rashba spin-orbit coupling. In general an electron in the system of interest can be scattered into the environment, travel a long distance, and then be scattered back to the system. Contributions from higher-order processes are important especially when the states of the system and that of the environment have similar energy,
{\it i.e.} when $\delta E$ is small. This is likely the main qualitative consideration influencing trends of effective Rashba spin-orbit coupling strengths across materials, and can therefore play 
a role in formulating strategies to optimize ferromagnetic chain topological superconductivity.   

The heuristic analysis explained above suggests that a Green's function (or the scattering) method \cite{williams_1982} 
might often be convenient in studying realistic systems, which we will also employ to study the spatial profile of the Majorana end modes in Sec. IVD. In this approach, the whole substrate is viewed as a scatterer and its influence on the electronic states of the chain it supports can be captured by a self-energy term $\Sigma_S$. The single-particle retarded Green's function of the chain is
\begin{eqnarray}\label{eq:Gr_chain}
G^r_{chain}(\omega)=[\omega+i\eta - H_{chain}-\Sigma_{S}(\omega)]^{-1},
\end{eqnarray} 
where $\eta$ is an infinitesimal real number. For example if we assume an infinite chain is along the $\hat{x}$ direction and a surface normal $\hat{z}$, 
\begin{eqnarray}\label{eq:SE0}
\Sigma_S(\omega,k_x)&=&h_t^\dag g_S h_t\\\nonumber
&=&\sum_{k_y} H^\dag_{t}(k_x,k_y) G_S(k_x,k_y) H_t(k_x,k_y)
\end{eqnarray} 
where $h_t$ is the hopping matrix between the chain and the substrate, and $g_S$ is the surface Green's function of the substrate which can be conveniently calculated using an iterative approach \cite{sancho_1985} when a tight-binding model of the substrate can be constructed. Note that we have used the convention that lower case letters stand for matrices of infinite dimension, while upper case letters refer to finite matrices diagonal in a momentum representation.  

If the substrate is metallic, $\Sigma_S$ will in general have a large non-hermitian 
contribution representing decay from the atomic chain into the substrate. 
Nonetheless, one can still crudely define the effective chain Hamiltonian including 
the substrate contribution as
 \begin{eqnarray}
H_{eff} \equiv H_{chain}+ \frac{1}{2}\left [\Sigma_{S}(\omega=0)+\Sigma_{S}^\dag(\omega=0) \right ].
\end{eqnarray}
If one is especially interested in the size of the induced Rashba spin-orbit coupling, it can be extracted from $\Sigma_{S}(\omega=0)$ as the net spin-flip hopping contribution that is odd in $k_x$,  similar to our definition of $t_R$ in Eq.~\ref{eq:tRdef}. However, the Rashba spin-orbit coupling will now be a matrix, have a nontrivial dependence on $k_x$, and sensitively depend on the relative positions of the chain and the lattice of the substrate. For model calculation purposes different approximations can be further made to obtain a manageable form of the Rashba spin-orbit coupling. One example following this approach is described in Ref.~\onlinecite{Nadj-Perge14}.

\section{Majorana States on a Transition Metal Ferromagnetic Chain}\label{sec:MS_chain}

\begin{figure}[h]
 \begin{center}
\includegraphics[width=2.5 in]{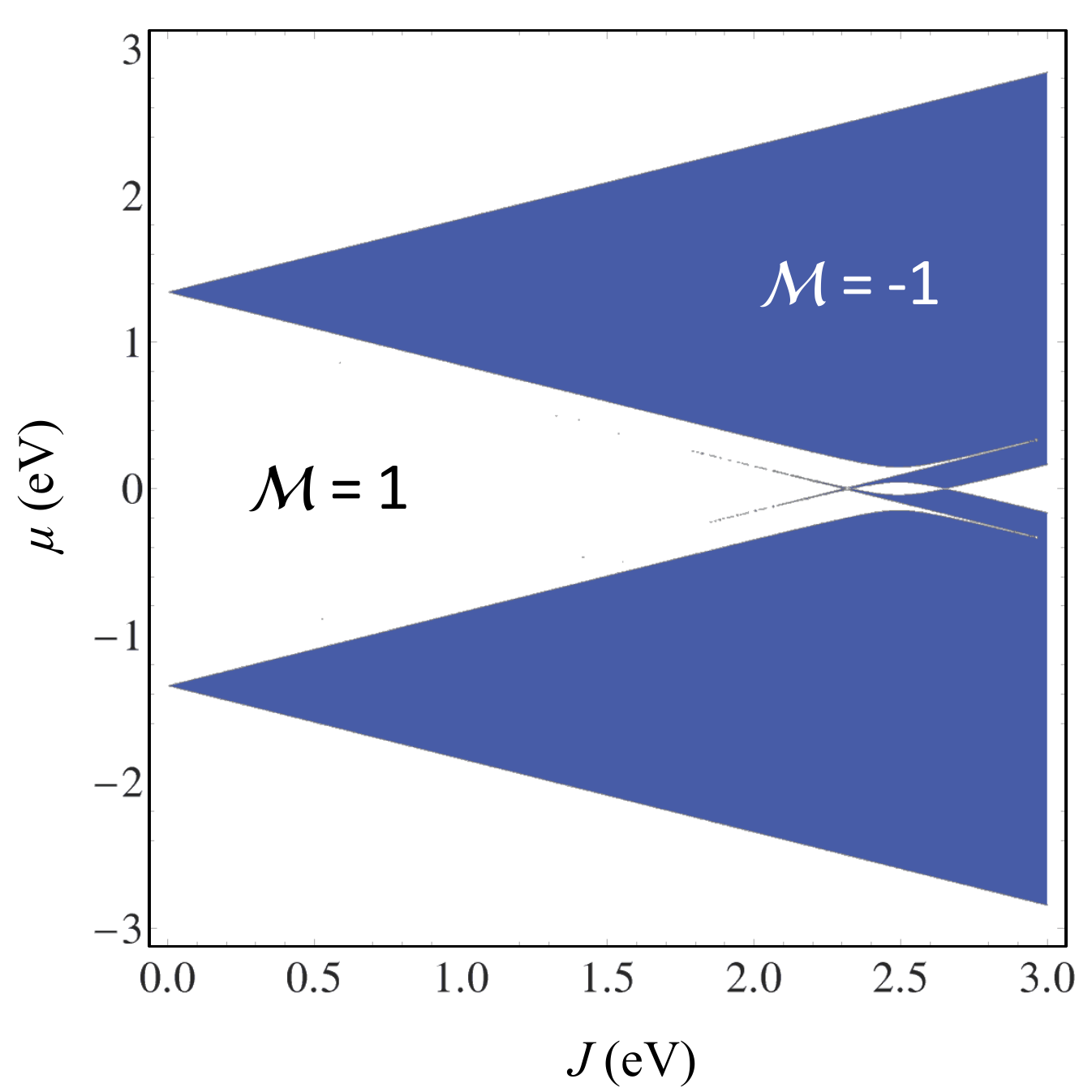}
 \end{center}
 \caption{Topological phase diagram for the $3d$ straight ferromagnetic chain model. Blue regions in chemical-potential {\it vs.} exchange coupling strength phase diagram have Majorana number $\mathcal{M}=-1$ while white regions have $\mathcal{M}=1$. This figure was constructed using the hopping parameters listed in Table I. When the exchange splitting is larger than the band width, the realistic case for transition metal chains, the gapped superconducting state is almost always topological.}\label{fig:5}
\end{figure}

Fig.~\ref{fig:5} illustrates where topological superconductivity occurs 
as a function of band filling and exchange splitting $J$ in straight transition metal chains. Note that in the physically realistic part of this phase diagram, where $J$ is comparable to or larger than the band width, the superconducting state is almost always topological for the reasons explained previously. 
This phase diagram has been determined by evaluating the Majorana number\cite{kitaev_2001} of an infinite chain, but is of course in agreement with the simple heuristic requirement that superconducting states are topological when the number of bands crossing the Fermi level in the absence of pairing is odd.

\begin{figure}[h]
 \begin{center}
\includegraphics[width=2.5 in]{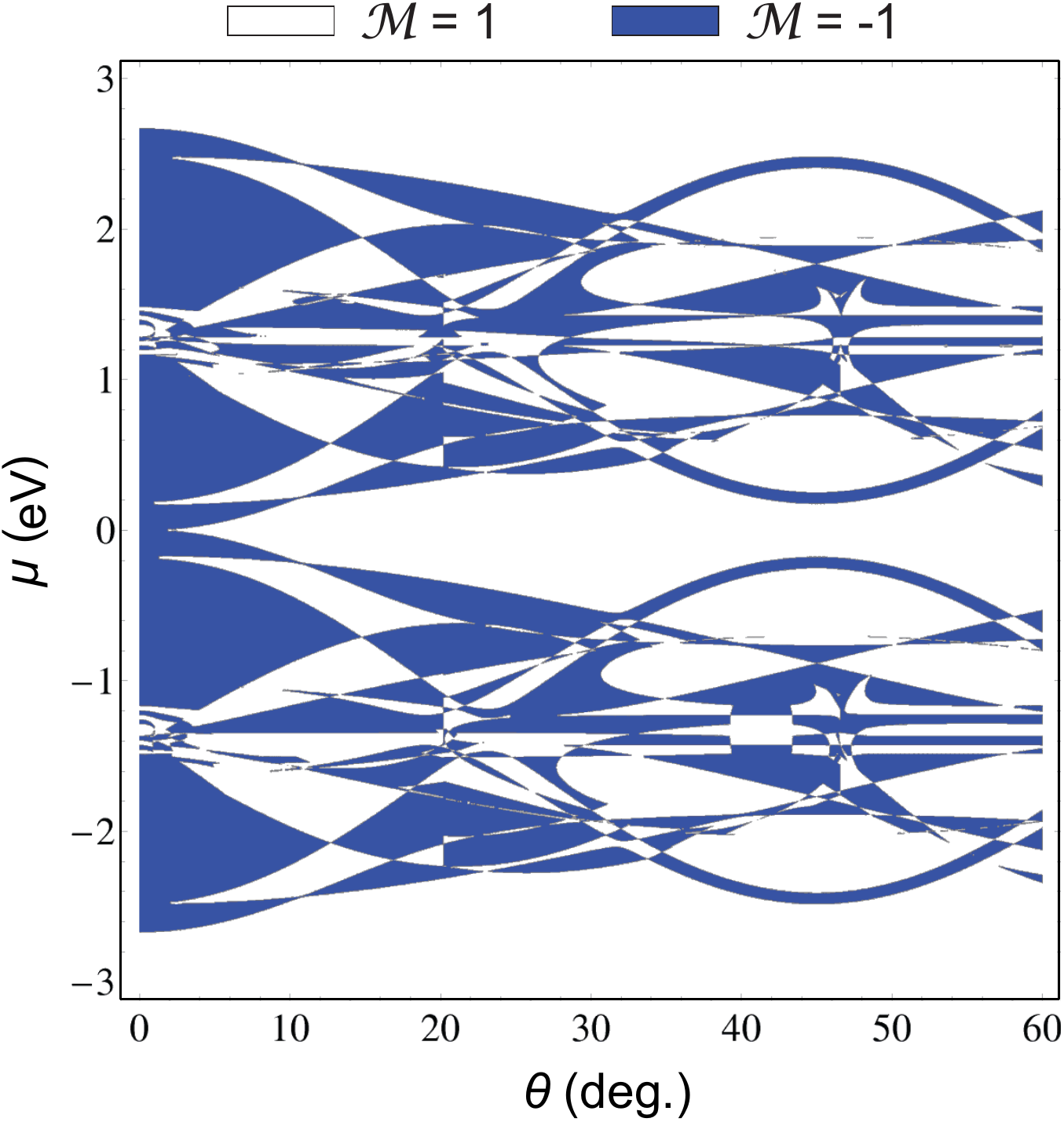}
 \end{center}
 \caption{(Color online) Majorana phase diagram of a zigzag chain with nearest neighbor hopping {\it vs.} chemical potential and 
 bond angle (cf. Fig.~\ref{fig:4}) at fixed exchange coupling $J=2.65$ eV. The blue and the white regions correspond to $\mathcal{M}=-1$ and $\mathcal{M}=1$, respectively.}\label{zigzagphasediagram}
\end{figure}

When placed on a superconducting substrate, transition metal atoms do not in general form straight chain structures. For example the structure formed by iron atoms in the chains studied in Ref.\onlinecite{Nadj-Perge14} consists of several rows of atoms. The structure in general will depend 
both on the details of the chemical bonding between transition metal and substrate atoms and on the growth protocol used to produce the chains.
Within single row structures, the straight chain can be generalized to zigzag chains in which the metal-metal-metal bonding angles alternate around $180^{\circ}$.  Angular momentum along the chain axis is no longer a good quantum number in zig-zag chains, and higher energy minority spin bands are no longer populated in pairs. As a result the topologically nontrivial regions in the phase diagram will in general shrink when the bonding angles deviate from $180^{\circ}$. 
This trend is illustrated in Fig.~\ref{zigzagphasediagram}, where we have fixed the exchange splitting at $J= 2.65 {\rm eV}$,
but varied the bond angle along a zigzag chain between $180^\circ$ (corresponding to a straight chain) and $120^\circ$.

\begin{figure}[h]
 \begin{center}
\includegraphics[width=3 in]{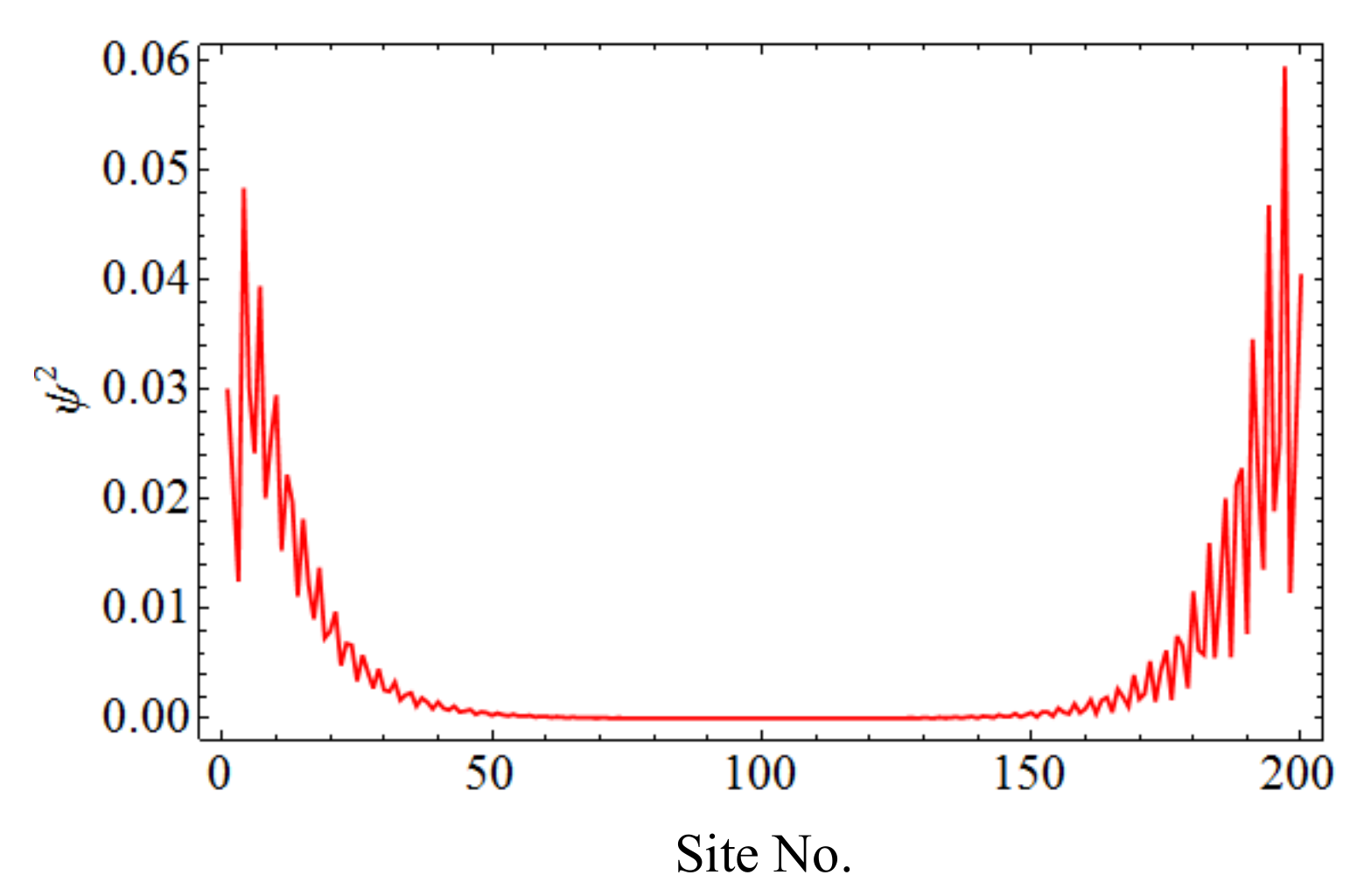}
\end{center}
\caption{Spatial distribution of one of the two Majorana states ($E\approx\pm 2\times 10^{-6}$ eV) for a finite chain with 
200 atoms. $\Delta=0.9$ eV, $t_R=0.1$ eV, so that the zero energy gap is $\sim 0.1$ eV in the infinite chain. }\label{fig:3}
\end{figure}

To confirm that zero energy Majorana modes exist in topologically nontrivial chains, we have also solved the BdG equations for finite length chains.
For example, when parameters are chosen so that the energy gap is $\sim 0.1$ eV in the infinite chain, we find two BdG eigenstates with $ |E| \approx 2\times 10^{-6}$ eV. Fig.~\ref{fig:3} demonstrates that these eigenstates are localized at the chain ends. 
The spatial extent of Majorana states in systems with realistic gap values will be discussed in detail in Sec. IVD.  

\begin{figure}[h]
 \begin{center}
\includegraphics[width=3 in]{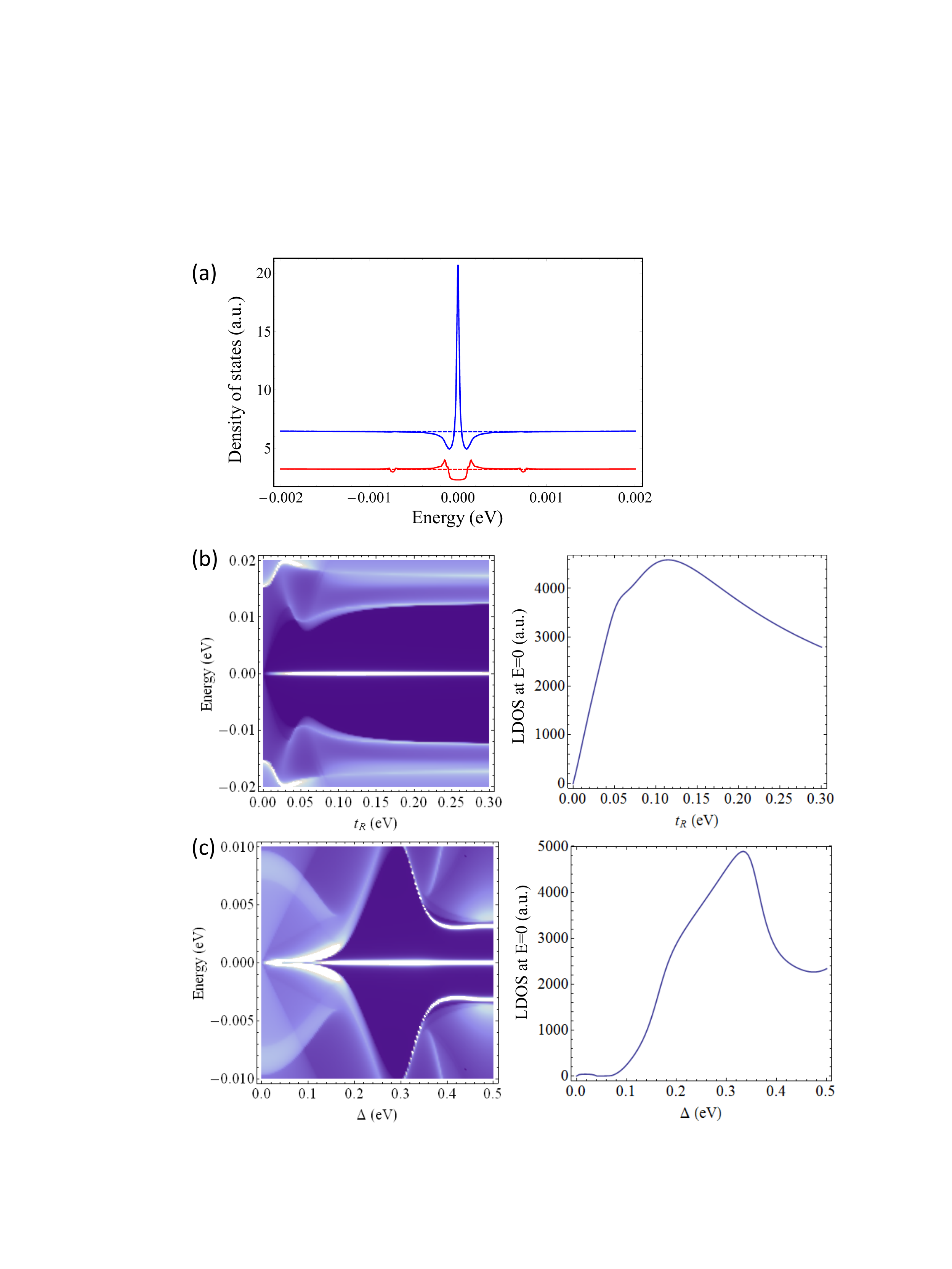}
 \end{center}
 \caption{ (a) Local density of states at the end of a semi-infinite chain (blue) and in the middle of an infinite chain (red). 
Both calculations were performed for chains with $\mathcal{M}=-1$. $\Delta=1.5$ meV, $t_R=0.1$ eV, $J=2.65$ eV, and $\mu = 1.3$ eV. (b) and (c), Non-monotonic dependence of the height of the zero energy peak with the parameters (b) $t_R$ and (c) $\Delta$. The left panel in each figure is the spectral function of the end Green's function in an energy window around zero energy, 
and the right panel is the zero energy value of the spectral function. In (b) $\Delta$ is fixed at 0.1 eV and in (c) $t_R$ is fixed at 0.1 eV. Values of the other parameters in (b) and (c) are the same as those in (a).}\label{fig:6}
\end{figure}

Given a model Hamiltonian the local density of states, which is closely related to STM $dI/dV$ data, can be conveniently 
calculated for infinite or semi-infinite chains using an iterative Green's function method.\cite{sancho_1985}  
In Fig.~\ref{fig:6} (a) we compare the local density of states at the end of a semi-infinite chain and in the 
middle of an infinite chain. Although both chains are topologically nontrivial 
and have the same parameter values, a zero-energy peak corresponding to the Majorana state appears only at the end of the semi-infinite chain.

The decay length of the Majorana states in the direction
toward the center of the chain is, roughly speaking, inversely proportional to the superconducting gap. 
However, for multi-orbital systems such a simple proportionality may not hold. 
To see this we point out that in our isolated chain model the decay length of the end states is proportional to the height of the
 local density of states peak. In Figs.~\ref{fig:6} (b) and (c) we plot the local density of states or spectral function at the end of a semi-infinite chain as a function of $s$-wave pairing $\Delta$ and Rashba spin-orbit coupling $t_R$, respectively. One can see that the height of the local density of states is neither monotonically proportional to these parameters, nor to the apparent superconducting gap of the chain. This behavior originates from the multi-orbital nature of the chain model, in which different bands may have different zero-energy splitting due to the same pairing potential, and each of them influences the decay length of the Majorana end state to some extent. We will discuss the length scale of the Majorana end states in detail with more realistic model calculations in Sec. IV.

\section{Hybrid system with an F\lowercase{e} Chain coupled to a P\lowercase{b}(110) Substrate}

\begin{figure}
  \centering
  \includegraphics[width=0.45\textwidth]{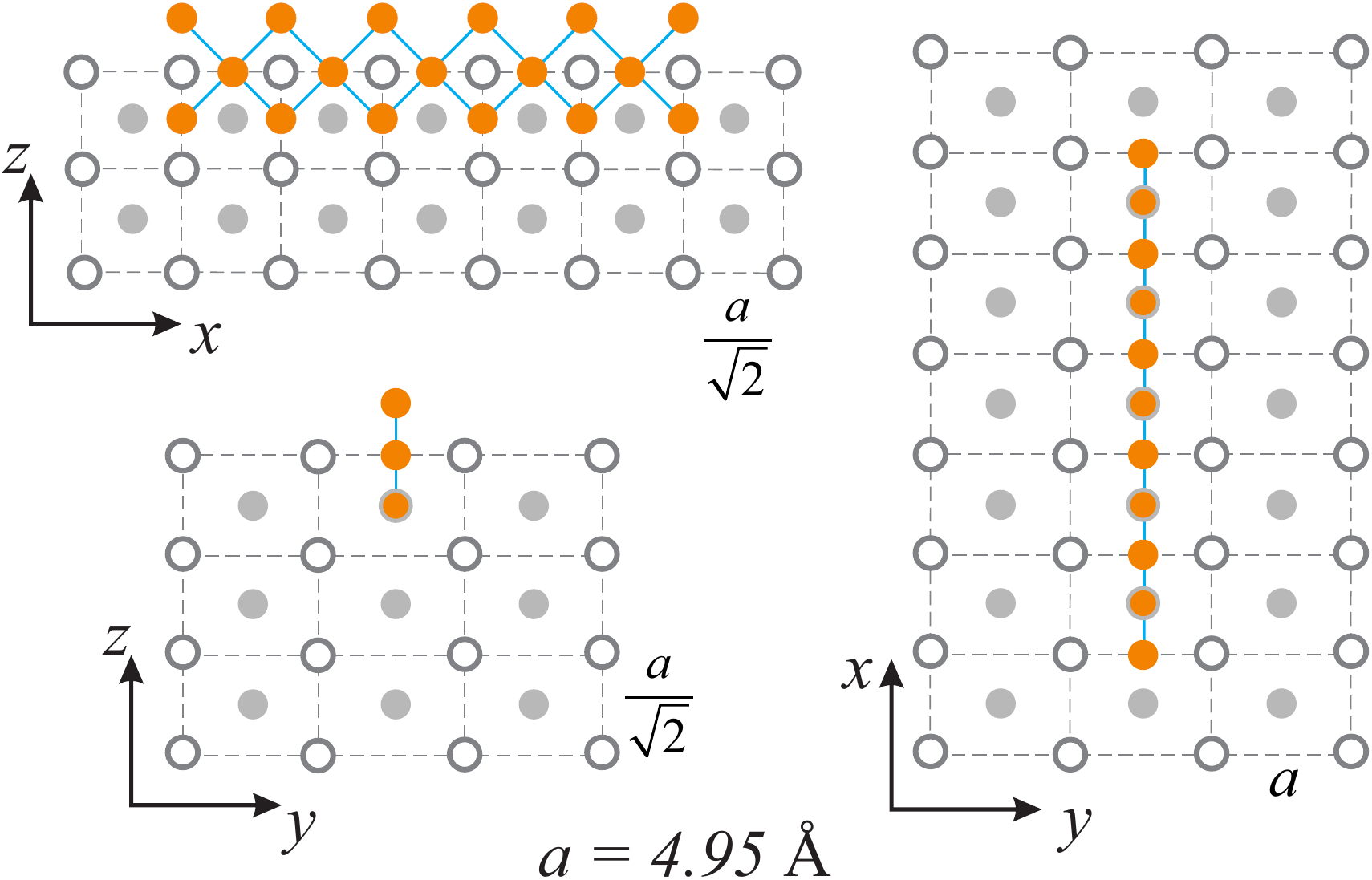}
  \caption{Geometry of the hybrid system: an Fe (orange) chain is embedded into the (110) surface of a bulk Pb (gray) superconducting substrate.}
  \label{fig:setup}
\end{figure}

The experimental system studied in Ref.~\onlinecite{Nadj-Perge14} is not purely one-dimensional. The Fe chain is embedded in a bulk Pb superconductor to 
which it is strongly hybridized.  The physics of this hybrid system is more complicated than that of a purely one-dimensional chain. This strong 
coupling between orbitals localized on the one-dimensional magnetic wire and those in the 
bulk superconductor has important consequences for many physical properties of the system, including the possible presence of multiple flavors of Majorana end modes, the spatial profile of the Majorana end modes and the presence of other in-gap bands along the wire known
as {\em Shiba} bands. These new elements distinguish the new platform \cite{Nadj-Perge14,pientka_2013} from 
other previously proposed Majorana host systems.

In this section we build a more realistic but still simplified model of our system by coupling the Fe tight-binding Hamiltonian to the Pb substrate through a tunneling term. This tunneling term induces both the Rashba-type spin-orbit coupling and the superconductivity in the Fe chain.  Both are essential ingredients for Majorana physics, but not native to Fe. The geometry of our model hybrid system, shown in Fig.~\ref{fig:setup}, is a commensurate version of the one obtained by comparing DFT calculations and experiments \cite{Nadj-Perge14}. Although it is most likely that the Fe atoms form triple chains in the samples investigated in Ref.~\onlinecite{Nadj-Perge14}, here we will first present conceptually important results by using linear chains as examples, to be consistent with the previous sections, and then discuss results for triple chains that are more relevant to experiments.
 
\subsection{Tight-binding Hamiltonian}

The tight-binding Hamiltonian for the hybrid system is
\begin{align}
  &H_{\text{hybrid}} = H_{\text{Fe}} + H_{\text{Pb}} + H_{\text{Fe-Pb}}, \label{eq:hybridham}\\
  &H_{\text{Fe}} = \sum_{\bm{r}} \bm{d}_{\bm{r}}^\dag \xi_{\text{Fe}}(\bm{r}) \bm{d}_{\bm{r}} + \sum_{\bm{r}_1\ne\bm{r}_2} \bm{d}_{\bm{r}_1}^\dag \tau_{\text{Fe}}(\bm{r}_1-\bm{r}_2) \bm{d}_{\bm{r}_2}, \\
  &H_{\text{Pb}} = \sum_{\bm{r}} \bm{c}_{\bm{r}}^\dag \xi_{\text{Pb}}(\bm{r}) \bm{c}_{\bm{r}} + \sum_{\bm{r}_1\ne\bm{r}_2} \bm{c}_{\bm{r}_1}^\dag \tau_{\text{Pb}}(\bm{r}_1-\bm{r}_2) \bm{c}_{\bm{r}_2} \nonumber \\
  &\qquad\quad + \sum_{\bm{r}} \bm{c}_{\bm{r}}^\dag \Delta(\bm{r}) (\bm{c}_{\bm{r}}^\dag)^T + h.c.\,, \\
  &H_{\text{Fe-Pb}} = \sum_{\bm{r}_1,\bm{r}_2} \bm{c}_{\bm{r}_1}^\dag \tau_{\text{Fe-Pb}}(\bm{r}_1-\bm{r}_2) \bm{d}_{\bm{r}_2} + h.c.\,.
\end{align}
Here $\bm{d}_{\bm{r}}^\dag$ ($\bm{c}_{\bm{r}}^\dag$) is the vector of electron creation operators for the Fe $3d$-orbitals (Pb $6p$-orbitals) and spins at site $\bm{r}$; $\xi$'s, $\tau$'s and $\Delta$ are matrices corresponding to normal on-site, hopping and conventional superconducting pairing terms, respectively. These matrices are explicitly given as follows
\begin{align}
  &\xi_{\text{Fe}}(\bm{r}) = \bigl\{[\epsilon_{\text{Fe}}(\bm{r})-\mu_{\text{Fe}}]s_0 -\bm{J}_{\text{Fe}}\cdot\bm{s}\bigr\} \otimes L_0^{(d)} \nonumber \\
  &\qquad\qquad+ \lambda_{\text{Fe}} \sum_{i=1}^{3} s_i\otimes L_i^{(d)}, \\
  &\xi_{\text{Pb}}(\bm{r}) = \epsilon_{\text{Pb}}(\bm{r})s_0 \otimes L_0^{(p)} + \lambda_{\text{Pb}} \sum_{i=1}^{3} s_i\otimes L_i^{(p)}, \\
  &\tau_{\text{Fe}}(\delta\bm{r}) = \sum_{\beta = \{\sigma,\pi,\delta\}} E_{dd\beta}(|\delta\bm{r}|)\, A_{dd\beta}(\delta\bm{r}/|\delta\bm{r}|) \otimes s_0, \\
  &\tau_{\text{Pb}}(\delta\bm{r}) = \sum_{\beta = \{\sigma,\pi\}} E_{pp\beta}(|\delta\bm{r}|)\, A_{pp\beta}(\delta\bm{r}/|\delta\bm{r}|) \otimes s_0, \\
  &\tau_{\text{Fe-Pb}}(\delta\bm{r}) = \hspace{-2mm}\sum_{\beta = \{\sigma,\pi\}} E_{pd\beta}(|\delta\bm{r}|)\, A_{pd\beta}(\delta\bm{r}/|\delta\bm{r}|) \otimes s_0, \\
  &\Delta(\bm{r})  = \Delta \,(i \sigma_2)\otimes L_0^{(p)}.
\end{align}
where $\epsilon_{\text{Fe}}$ and $\epsilon_{\text{Pb}}$ are on-site energies, $\mu_{\text{Fe}}$ is the chemical potential, $\bm{J}_{\text{Fe}}$ is the magnetization vector in Fe, $\lambda_{\text{Fe}}$ and $\lambda_{\text{Pb}}$ are atomic spin-orbit coupling energies, $E_{dd\beta}$, $E_{pp\beta}$ and $E_{pd\beta}$ are the Slater-Koster bond integrals that depend on the types of bond ($\beta$) and the distance between atoms ($|\delta\bm{r}|$),  $A_{dd\beta}$, $A_{pp\beta}$ and $A_{pd\beta}$ are the real coefficient matrices of Slater-Koster integrals in the cubic harmonic basis and are dependent only on the relative angle between atoms, \cite{slater_1954} $\Delta$ is the (real) $s$-wave pairing potential,  $\bm{s}$ and $\bm{L}$ (the superscripts indicating the type of the orbitals) are spin and orbital angular momentum operators with $s_0$ and $L_0$ the corresponding identity matrices. We use the convention $\bm{s} = \frac{1}{2}\bm{\sigma}$ where $\bm{\sigma}$ is the vector of Pauli matrices, and
\begin{align}
  &L_1^{(p)} =
  \begin{pmatrix}
    0 & 0 & 0 \\
    0 & 0 & i \\
    0 & -i & 0
  \end{pmatrix},\quad
  L_2^{(p)} =
  \begin{pmatrix}
    0 & 0 & i \\
    0 & 0 & 0 \\
    -i & 0 & 0
  \end{pmatrix},\nonumber\\
  &L_3^{(p)} =
  \begin{pmatrix}
    0 & i & 0 \\
    -i & 0 & 0 \\
    0 & 0 & 0
  \end{pmatrix},\\
  &L_1^{(d)} =
  \begin{pmatrix}
    0 & 0 & i & 0 & 0 \\
    0 & 0 & 0 & i & \sqrt{3}i \\
    -i & 0 & 0 & 0 & 0 \\
    0 & -i & 0 & 0 & 0 \\
    0 & -\sqrt{3}i & 0 & 0 & 0
  \end{pmatrix},\nonumber\\
  &L_2^{(d)} =
  \begin{pmatrix}
    0 & i & 0 & 0 & 0 \\
    -i & 0 & 0 & 0 & 0 \\
    0 & 0 & 0 & -i & \sqrt{3}i \\
    0 & 0 & i & 0 & 0 \\
    0 & 0 & -\sqrt{3}i & 0 & 0
  \end{pmatrix},\nonumber\\
  &L_3^{(d)} =
  \begin{pmatrix}
    0 & 0 & 0 & -2i & 0 \\
    0 & 0 & -i & 0 & 0 \\
    0 & i & 0 & 0 & 0 \\
    2i & 0 & 0 & 0 & 0 \\
    0 & 0 & 0 & 0 & 0
  \end{pmatrix}.
\end{align}

The above Hamiltonian has a very general form. When the Fe atoms lie in a mirror plane of the Pb lattice, as in Fig.~\ref{fig:setup} for example, and when $\epsilon_{\text{Fe}}(\bm{r})$, $\epsilon_{\text{Pb}}(\bm{r})$ and $\bm{J}_{\text{Fe}}$ are all symmetric with respect to the same mirror plane, the Hamiltonian satisfies an anti-unitary symmetry that combines mirror and time-reversal operations. Assuming the $xz$ plane to be the mirror plane, the mirror and time reversal symmetry operators, in the cubic harmonic basis for orbitals, are given by
\begin{align}
  &M_{xz} = (-1)^l \exp(-i\pi L_2)\otimes\exp(-i\pi s_2)\mathcal{M}(y\rightarrow-y), \\
  &T = L_0\otimes\exp(-i\pi s_2)\mathcal{K},
\end{align}
where $l$ is the orbital angular momentum quantum number ($l=1$ for $p$-orbitals; $l=2$ for $d$-orbitals), $\mathcal{M}(y\rightarrow-y)$ stands for the real-space mirror reflection with respect to the $xz$ plane, and $\mathcal{K}$ is the complex conjugate operator. The invariance of the Hamiltonian under the combined symmetry $M_T = M_{xz}T$ can be broken down to the following invariance relations:
\begin{align}
  &M_T \,\xi_{\text{Fe}}(x,y,z)\, M_T^{-1} = \xi_{\text{Fe}}(x,-y,z),\\
  &M_{xz} \,\xi_{\text{Pb}}(x,y,z)\, M_{xz}^{-1} = \xi_{\text{Pb}}(x,-y,z),\\
  &T \,\xi_{\text{Pb}}(x,y,z)\, T^{-1} = \xi_{\text{Pb}}(x,y,z),\\
  &M_{xz} \,\tau(\delta x,\delta y,\delta z)\, M_{xz}^{-1} = \tau(\delta x,-\delta y,\delta z),\\
  &T \,\tau(\delta x,\delta y,\delta z)\, T^{-1} = \tau(\delta x,\delta y,\delta z),
\end{align}
where $\tau$ stands for each of $\tau_{\text{Fe}}$, $\tau_{\text{Pb}}$ and $\tau_{\text{Fe-Pb}}$. In addition, the invariance of the superconducting pairing term under $M_T$ is trivially satisfied. In the Nambu basis, the BdG Hamiltonian satisfies the particle-hole (charge conjugation) symmetry given by
\begin{align}
  C = L_0\otimes\exp(-i\pi s_2)\otimes(i \rho_2)\mathcal{K},
\end{align}
where $\bm{\rho}$ is the vector of Pauli matrices for the particle-hole degree of freedom. The combination of $M_T$ and $C$ results in a chiral symmetry
\begin{align}
  U_{\chi} = M_T C = M_{xz}\otimes(-i \rho_2),
\end{align}
which is unitary and transforms the Hamiltonian as $U_{\chi} H_{\text{hybrid}} U_{\chi}^{-1}=-H_{\text{hybrid}}$. The implications of these symmetries will be analyzed in detail in Sec.~\ref{ssec:sym}.

Most of our following results are obtained by performing exact diagonalizations of the above Hamiltonian. In order to maintain a limited yet realistic parameter set, we further assume
\begin{align}
  &\epsilon_{\text{Fe}}(\bm{r}) = \epsilon_{\text{Fe}},\quad \epsilon_{\text{Pb}}(\bm{r}) = \epsilon_{\text{Pb}}, \\
  &\bm{J}_{\text{Fe}} = (0,0,J_{\text{Fe}}), \\
  &E_{dd\beta}(|\delta\bm{r}|) = V_{dd\beta}\,(r_0/|\delta\bm{r}|)^{n_{dd\beta}}\quad (|\delta\bm{r}|\le a/\sqrt{2}), \label{eq:Edd} \\
  &E_{pp\beta}(|\delta\bm{r}|) = V_{pp\beta}^1\quad \text{if}\quad |\delta\bm{r}|= a/\sqrt{2}, \\
  &E_{pp\beta}(|\delta\bm{r}|) = V_{pp\beta}^2\quad \text{if}\quad |\delta\bm{r}|= a, \\
  &E_{pd\beta}(|\delta\bm{r}|) = V_{pd\beta}\,(\frac{a}{\sqrt{8}}/|\delta\bm{r}|)^{n_{pd\beta}}\quad (|\delta\bm{r}|\le \sqrt{\frac{3}{8}}a), \label{eq:Epd}
\end{align}
where $r_0=2.383\text{\AA}$ is the nearest neighbor distance in bulk Fe(bcc), and $a = 4.95\text{\AA}$ is the lattice constant of bulk Pb(fcc). Eqs.~(\ref{eq:Edd}-\ref{eq:Epd}) imply that in all types of hopping terms we include up to the second nearest neighbors (cf. Fig.~\ref{fig:setup}). We list all the parameters, except for $\epsilon_{\text{Fe}}$ and $\Delta$, and their references (if applicable) in Table~\ref{tab:allpars}. Since $\epsilon_{\text{Fe}}$ and $\mu_{\text{Fe}}$ are not actually independent parameters in the model, $\epsilon_{\text{Fe}}$ will be chosen in the linear Fe chain case such that $\mu_{\text{Fe}}=0$ corresponds to the center of the minority-spin band, and in the triple Fe chain case according to experiment.\cite{Nadj-Perge14}

\begin{table}
\caption{Parameters for the tight-binding Hamiltonian. The undetermined parameters, $\mu_{\text{Fe}}$ and $V_{pd\pi}$ are variables in the simulations.}
\centering
$\,$\\
\begin{tabular}{cccccccc}
\hline\hline
Parameters & Ref. & Value & \vline&\vline & Parameters & Ref. & Value \\\hline
$\mu_{\text{Fe}}$ & & ? & \vline&\vline & $\epsilon_{\text{Pb}}$ & \onlinecite{wurde_1994} & 0.97 eV \\
$\lambda_{\text{Fe}}$ & \onlinecite{lambdaFe} & 0.06 eV & \vline&\vline & $\lambda_{\text{Pb}}$ & \onlinecite{wurde_1994} & 0.665 eV \\
$V_{dd\sigma}$ & \onlinecite{zhong_1993} & -0.6702 eV & \vline&\vline & $V_{pp\sigma}^1$ & \onlinecite{wurde_1994} & 1.134 eV \\
$V_{dd\pi}$ & \onlinecite{zhong_1993} & 0.5760 eV & \vline&\vline & $V_{pp\pi}^1$ & \onlinecite{wurde_1994} & 0.080 eV \\
$V_{dd\delta}$ & \onlinecite{zhong_1993} & -0.1445 eV & \vline&\vline & $V_{pp\sigma}^2$ & \onlinecite{wurde_1994} & 0.146 eV \\
$n_{dd\sigma}$ & \onlinecite{zhong_1993} & 3 & \vline&\vline & $V_{pp\pi}^2$ & \onlinecite{wurde_1994} & 0 \\
$n_{dd\pi}$ & \onlinecite{zhong_1993} & 4 & \vline&\vline & $V_{pd\sigma}/V_{pd\pi}$ & \onlinecite{harrison_1980} & -2.17 \\
$n_{dd\delta}$ & \onlinecite{zhong_1993} & 4 & \vline&\vline & $n_{pd\sigma}$, $n_{pd\pi}$ & \onlinecite{andersen_1978} & 4 \\
$J_{\text{Fe}}$ & DFT & 2.5 eV & \vline&\vline & $V_{pd\pi}$ & & ? \\
\hline\hline
 \end{tabular}\label{tab:allpars}
\end{table}

\subsection{Multi-Majorana Chains Protected by a Magnetic Symmetry}\label{ssec:sym}

In this section we investigate a new symmetry that could be present in our systems, and which permits the presence of multiple Majorana fermions at the 
end of the chain. It was first shown in the theoretical part of the supplementary material of Ref.~\onlinecite{Nadj-Perge14} that in certain cases, multiple 
Majorana zero modes can appear at the end of the Fe chain. We here explain those results. The conditions needed are: the chain must be perfectly straight within a mirror plane of the Pb substrate; the magnetic moment of the iron must have no component perpendicular to the chain and 
parallel to the Pb surface; the Pb substrate must be disorder free. 
In this case we show that a magnetic symmetry first proposed in Ref.~\onlinecite{fang_2014} can stabilize an integer number of Majorana fermions
at the end of the Fe chain.

\begin{figure}
  \centering
  \includegraphics[width=0.5\textwidth]{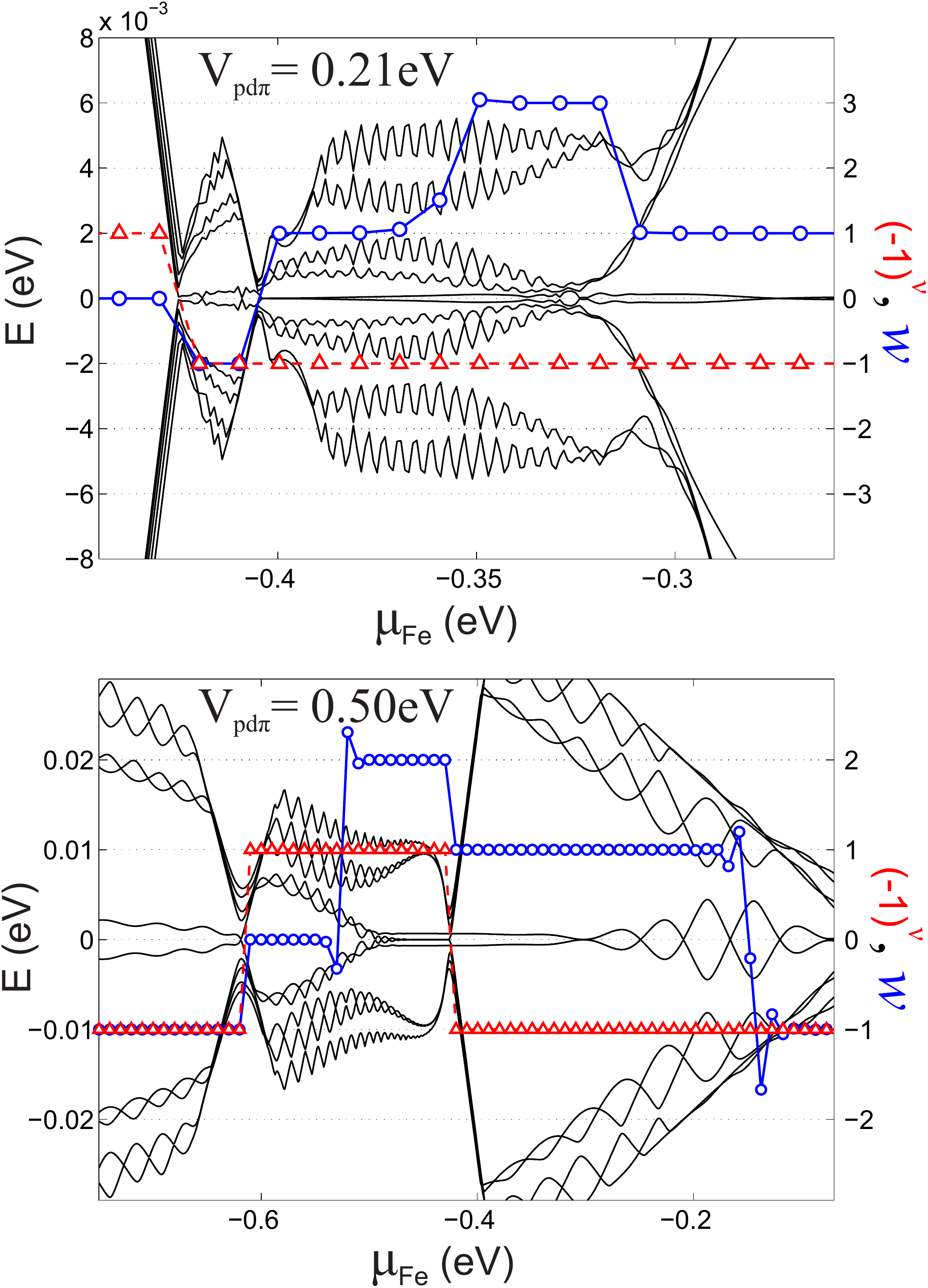}
  \caption{Typical low energy spectra of a finite-size hybrid system (left axis) with a linear Fe chain along with the corresponding values of both topological invariants (right axis), the Majorana number $\nu$ and the winding number $w$, computed from the bulk Hamiltonian. In these calculations the 
Fe chain is 120 unit cells long, $\Delta = 0.1$ eV, and the finite size of the Pb substrate is 21 unit cells in the $y$-direction, 1 in the $z$-direction and 140 in the $x$-direction.  The parameter $\mu_{Fe}$ which is varied specifies the band line up between the Fe and Pb states as described 
in the text.  Results are presented for two different strengths of the Pb/Fe hybridization parameter $V_{pd\pi}$.}  
\label{fig:sp}
\end{figure}

For a straight Fe chain along the $x$ direction, on an infinite $xy$ Pb surface with the $z$ direction perpendicular to the surface, 
and in the \emph{absence} of any magnetism in the chain, the $xz$ plane is a mirror plane. We call the mirror operator along that plane 
$M_{xz}$ with the properties:
\begin{equation}
[M_{xz}, H]=0, \;\;\; M_{xz}^2=-1
\end{equation}
where $H$ is the superconducting Pb and \emph{non-magnetic} Fe hybrid structure Hamiltonian operator. Without magnetism, the system is also time-reversal invariant with a spinful time-reversal operator $T$, $T^2=-1$. 

We now add magnetism in the system, on the Fe chain. A magnetic moment breaks time-reversal symmetry which generically also breaks 
the $M_{xz}$ mirror symmetry.  Only a magnetic moment polarized along the $y$ direction does not break $M_{xz}$, but still breaks $T$. 
However, if localized in the $xz$ plane, the magnetic moment is still invariant under the combination of mirror and time-reversal, a magnetic symmetry $M_T= M_{xz} T$. This is indeed true as each of the operations flips the magnetic moment so that their combination leaves it untouched. 
This magnetic symmetry was considered first in Ref.~\onlinecite{fang_2014}, in a different context, where it was shown that it stabilizes an integer $Z$ number of Majoranas in the vortex core of a crystalline topological insulator.  We here repeat the argument to show this symmetry.

The magnetic symmetry $M_T$ has the properties (since $[M_{xz}, T]=0$):
\begin{equation}
[M_T, H_{\text{hybrid}}]=0, \;\;\; M_T^2 = M_{xz}^2 T^2= 1, \;\;\; [M_T, C]=0
\end{equation}  where $C$ is the charge conjugation operator. Hence $M_T$ acts like spinless time-reversal (squares to 1 and it is antiunitary), and it can stabilize multiple Majoranas at the edge because any mass terms $i \gamma_a \gamma_b$ between any Majorana are not allowed due to the $i$ which breaks $M_T$ because of the complex conjugation.

In terms of topological classifications, \cite{ryu2010} our system falls into the BDI symmetry class because of the presence of both $M_T$ and $C$ symmetries, and hence a chiral symmetry $U_{\chi}=M_T C$. The bulk of the hybrid system, which is effectively 1D inside the superconducting gap of the Pb substrate, can be classified by a winding number
\begin{align}
  w = i\int_0^{2\pi} \frac{dk}{2\pi} \,\text{Tr}[h(k)^{-1}\partial_k h(k)],
\end{align}
where $h(k)$ is defined such that the Bloch Hamiltonian $H_{\text{hybrid}}(k)$ is brought to the following form by the eigenstates of charge conjugation operator
\begin{align}
  &V_{\chi}^\dagger H_{\text{hybrid}}(k)  V_{\chi} =
  \begin{pmatrix}
    0 & h(k) \\
    h(k)^\dagger & 0
  \end{pmatrix},\\
  &V_{\chi}^\dagger U_{\chi}  V_{\chi} =
  \begin{pmatrix}
    \mathds{1} & 0 \\
    0 & -\mathds{1}
  \end{pmatrix}.
\end{align}
Furthermore, the Majorana number $\nu$, defined and investigated in the previous sections (where $\mathcal{M}=(-1)^\nu$ has been used), is related to the winding number by $\nu = w\mod 2$.

\begin{widetext}
\begin{figure*}
  \centering
  \includegraphics[width=0.95\textwidth]{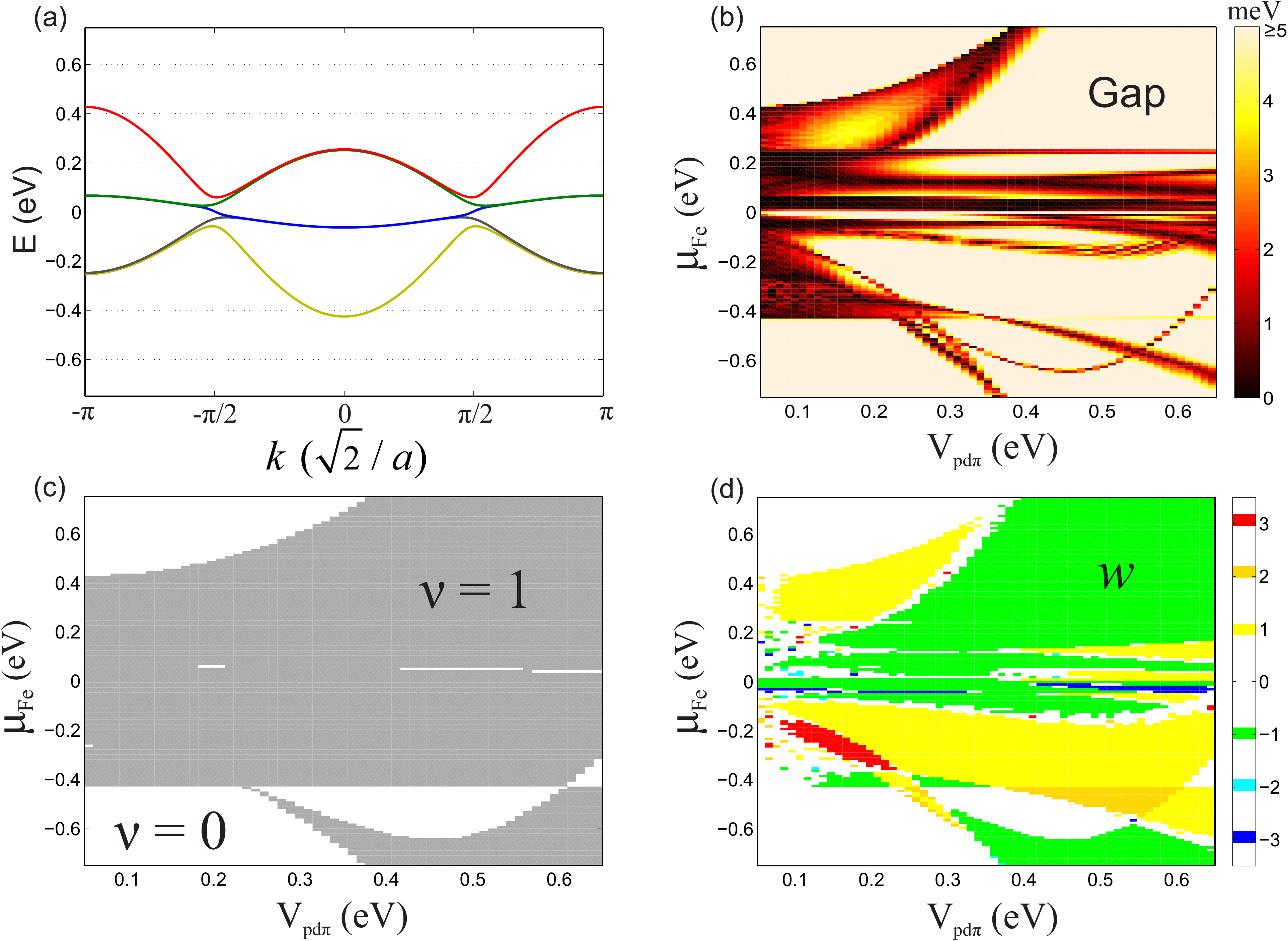}
  \caption{Phase diagrams of a hybrid structure with a linear Fe chain and a 2D superconducting Pb substrate. Panel (a) shows the band structure (for the minority band only) of the Fe chain when it is suspended. Panels (b), (c) and (d) are all plotted as a function of $\mu_{\text{Fe}}$ and $V_{pd\pi}$, showing the gaps of the hybrid structure, the Majorana number $\nu$, and the topological invariant (winding number) $w$, respectively. 
The parameter $V_{pd\pi}$ signifies the strength of the hybridization. The calculations of $w$ suffer significantly from numerical errors when the system gap is small. In order to reduce these errors, we set $\Delta = 0.1$ eV in the present calculations.  When the errors of $w$ is insignificant, we find $\nu = w\mod 2$, as expected.
The size of the Pb substrate used here is 21 unit cells in the $y$-direction and 1 in the $z$-direction (infinite in the $x$-direction).}
  \label{fig:pd}
\end{figure*}
\end{widetext}

We now exemplify the above reasoning for the specific model for Fe chains on the surface of Pb explained above. 
As presented previously, the hybrid Hamiltonian~\eqref{eq:hybridham} is invariant under the magnetic symmetry $M_T$. We therefore expect multiple pairs of Majoranas appearing at the end of the chain. This is indeed confirmed by diagonalizing the Hamiltonian for a finite-size hybrid system. Several such examples are shown in Fig.~\ref{fig:sp}. In addition, we show phase diagrams of such a hybrid structure in Fig.~\ref{fig:pd}, where in particular a phase diagram of the winding numbers (the topological invariant for a Hamiltonian exhibiting $M_T$) as a function of $\mu_{\text{Fe}}$ and $V_{pd\pi}$ is shown in Fig.~\ref{fig:pd}(c).

This new idea could also potentially allow us to experimentally investigate  interaction effects in Majorana fermions. 
While the multiple integer Majorana classification is noninteracting, we expect that interactions will lift the degeneracy of $8$ Majoranas providing a $Z \rightarrow Z_8$ classification. With significant experimental effort, this could be potentially tested in the future.

\subsection{Phase Diagram of a Triple Fe Chain Hybridized with a 2D Pb Substrate}

\begin{figure}
  \centering
  \includegraphics[width=0.47\textwidth]{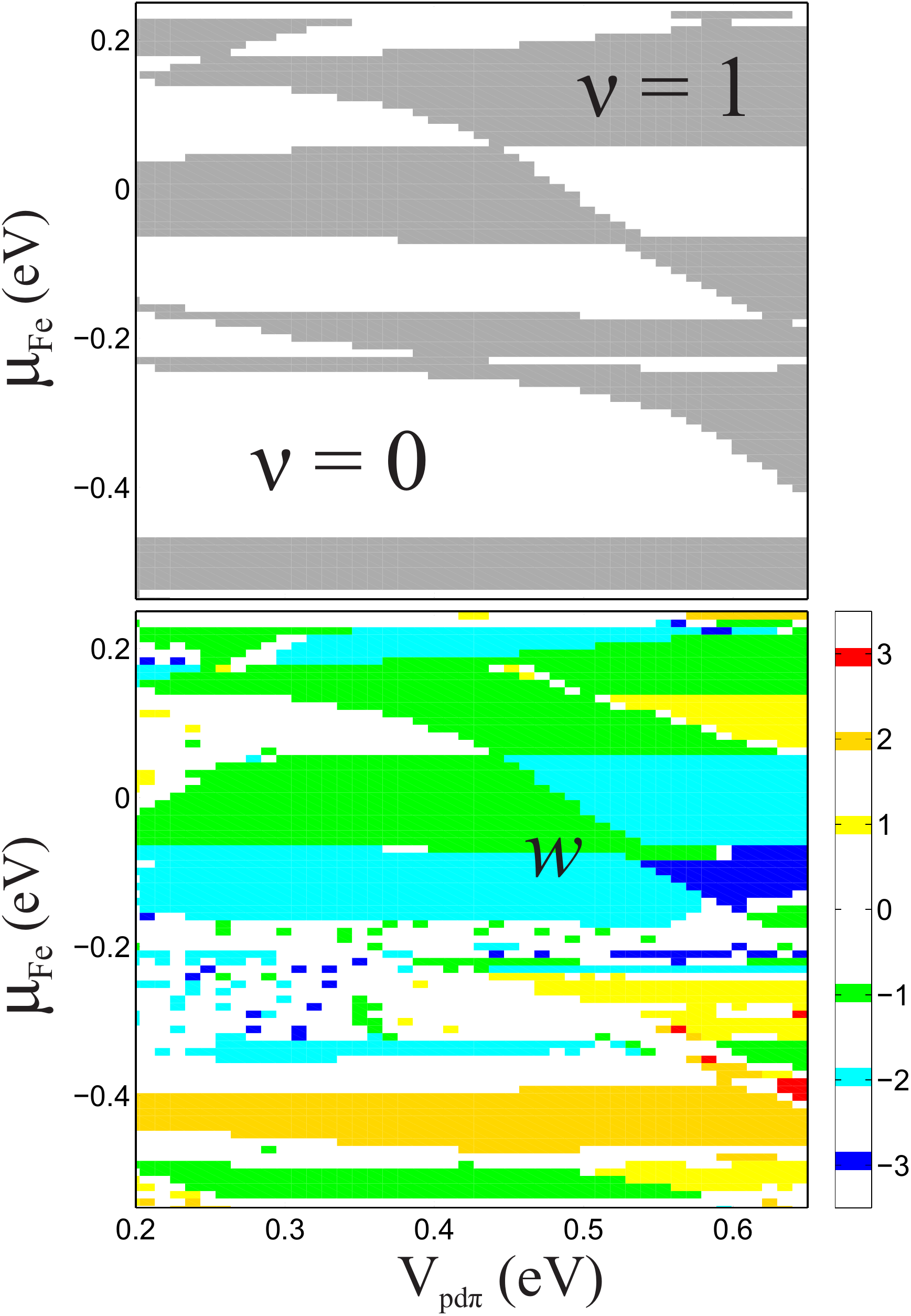}
  \caption{Phase diagrams of a hybrid structure with a triple Fe chain coupled to a 2D superconducting Pb substrate. The upper panel shows the phase diagram in terms of the Majorana number $\nu$ while the lower panel shows the phase diagram in terms of the the winding number $w$. 
In these calculations, $\Delta = 0.1$ eV, and the size of the Pb substrate is 15 unit cells in the $y$-direction and 1 in the 
$z$-direction (infinite with periodic boundary conditions in the $x$-direction).}
  \label{fig:pd3}
\end{figure}

We now discuss the hybrid structure with a triple Fe chain, which is more relevant to existing experiments presented in Ref.~\onlinecite{Nadj-Perge14}. In this case the band structure of a suspended Fe chain is significantly more complicated than that of the linear chain shown in Fig.~\ref{fig:2} and Fig.~\ref{fig:pd}(a). STM measurements suggest that the Fermi energy in the triple chain is likely to lie between two sets of narrow bands predicted by
a chain band structure calculation \cite{Nadj-Perge14}. We will refer to this energy as the $\mu_{\text{Fe}}=0$ point in the following presentations.
In Fig.~\ref{fig:pd3} we show two phase diagrams in the $\{\mu_{\text{Fe}},V_{pd\pi}\}$ space, in terms of the Majorana numbers and the winding numbers, respectively. Remarkably, but not surprisingly, although the phase diagram in terms of the Majorana numbers contains almost equal areas for $\nu=1$ and $\nu=0$ phases, because of the larger number of degeneracies lifted in a triple chain compared with a linear chain, the phase diagram in terms of the winding numbers is still dominated by topologically nontrivial ($w\ne 0$) phases. This implies that Majorana end modes are almost certainly present in the current hybrid structure. To further identify the actual number of Majorana modes in experiments will be interesting but challenging.

\subsection{Spatial Extent of the Majorana Fermions}

\begin{figure}
  \centering
  \includegraphics[width=0.47\textwidth]{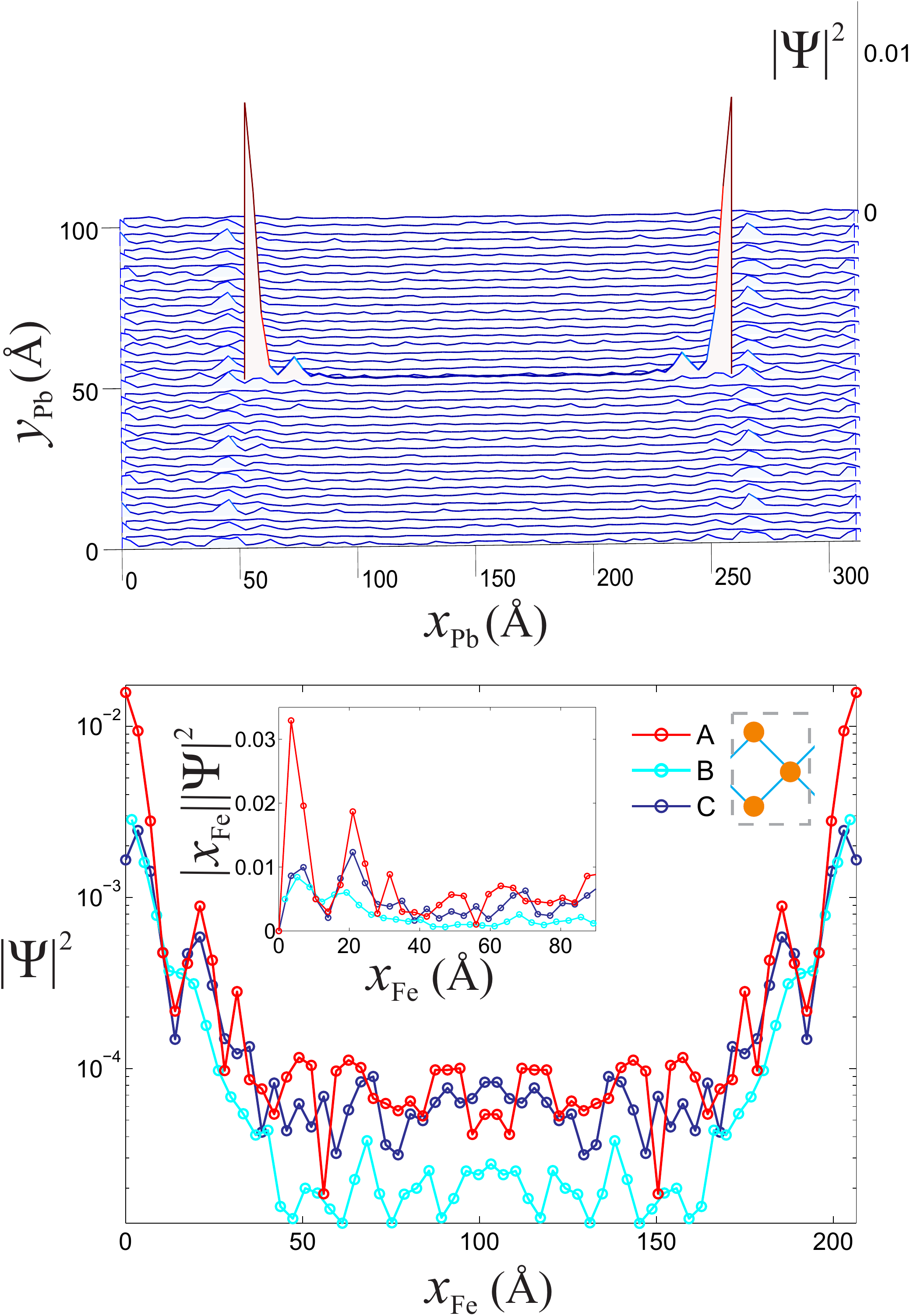}
  \caption{A representative eigenstate wavefunction of a finite-size hybrid structure composed of a 60-unit-cell-long triple Fe chain and a Pb substrate of size 90$\times$21$\times$2 unit cells. The upper panel shows the wavefunction amplitude on both the Fe chain and the Pb substrate (amplitudes for overlap points in the $xy$ plane are summed up); the lower panel shows details of the wavefunction on the Fe chain separately for each sublattice (see text). The energy corresponding to this eigenstate is 0.09 meV (see the highlighted point in Fig.~\ref{fig:sp_comp} upper panel); the ratio of the total weight of this wavefunction on the Pb substrate to that on the Fe chain is approximately 9.3 because of the strong hybridization and the significantly larger size of the substrate. Periodical boundary conditions for the substrate are adopted in both the $x$ and the $y$ directions in order to stabilize our numerical calculations, which leads to relatively strong coupling between the end states across the boundaries (see the enhanced wavefunction amplitudes in the substrate across the boundaries). The pairing parameter in Pb has a realistic value $\Delta = 1.3$ meV; the coupling parameter in this specific example is $V_{pd\pi} = 0.65$ eV.}
  \label{fig:mf_wf}
\end{figure}

\begin{figure}
  \centering
  \includegraphics[width=0.45\textwidth]{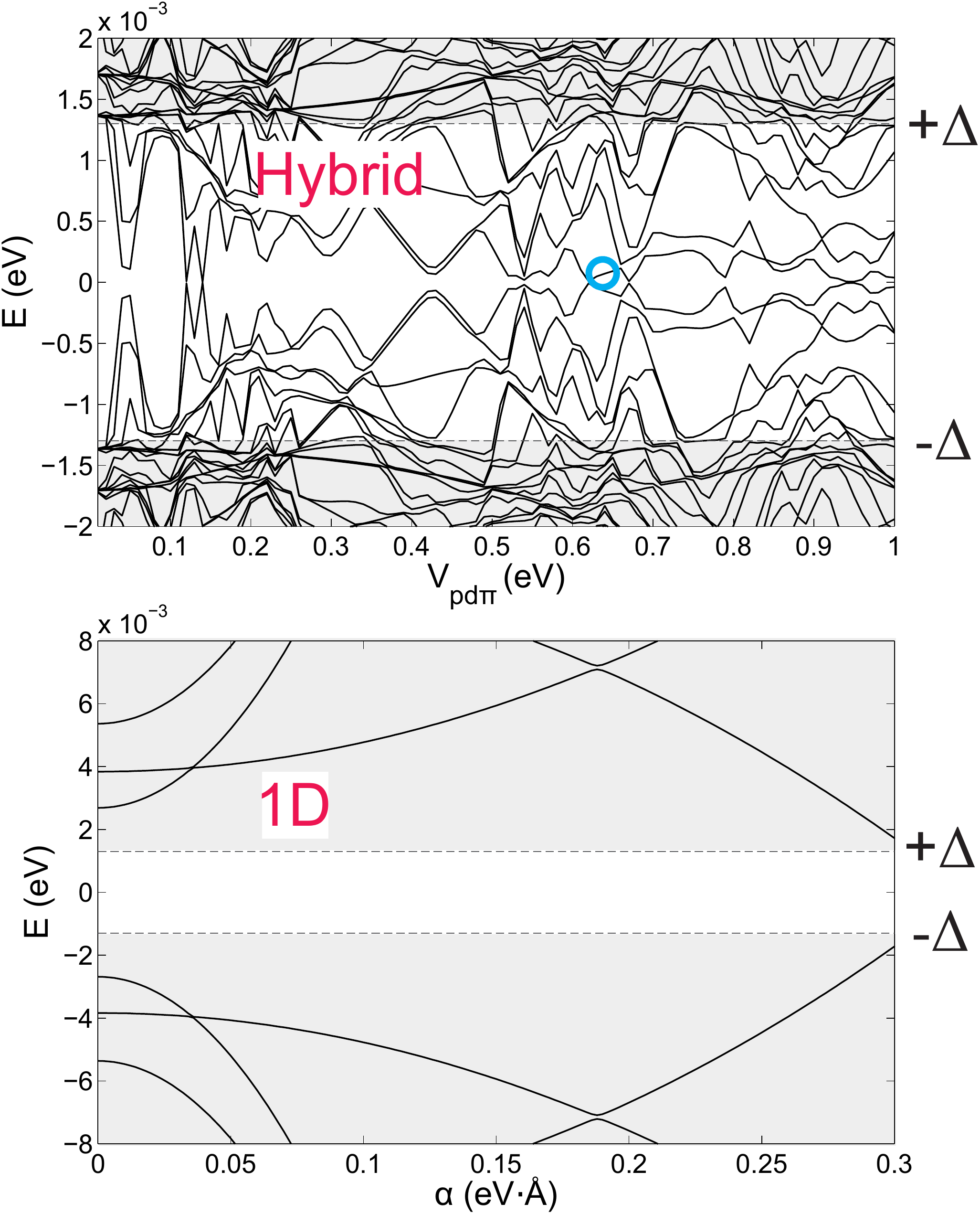}
  \caption{Comparison between the low-energy spectrum in the hybrid system (upper panel) and that in a suspended triple Fe chain with artificially added Rashba spin-orbit coupling and spin-singlet pairing terms (lower panel). In the upper panel, the spectrum is plotted as a function of the coupling between Fe and Pb atoms; in the lower panel, the spectrum is plotted as a function of the Rashba spin-orbit coupling strength ($\alpha$) in a realistic range. The two models have Fe chains of the same length (about 210 \AA), and the same pairing parameter ($\Delta = 1.3$ meV). The small circle in the upper panel highlights the eigenstate that is plotted in Fig.~\ref{fig:mf_wf}.}
  \label{fig:sp_comp}
\end{figure}

The hybrid nature of our setup is  most evident in its influence on the spatial extent of the Majorana end modes. 
In a purely one-dimensional system, it is well known that the Majorana modes are exponentially localized at the end of the chain as $\exp(-r/L)$ . 
The localization length $L \sim t/\Delta_p$ (in units of the chain lattice constant), equal to the coherence length of the effective $p$-wave superconductor.
 In our system, the $p$-wave wire coherence length inferred from the measured gap near the middle of the chain
 is very large ($L\sim 10^4$ unit cells) because the proximity induced gap $\Delta_p$ is an order of magnitude smaller than that of a bulk Pb 
superconductor $\Delta$, of the order $\Delta_p \sim \Delta E_{SO}/J$. 
If the system were purely one-dimensional, the localization length of the Majorana end states would have been much larger than the length of the chain (typically $\lesssim 10^2$ lattice constants) and no zero bias anomaly would have been observed near the chain ends.
However, when the one-dimensional chain is embedded in the higher-dimensional superconductor, the spatial profile of the Majorana end mode is predicted to acquire an additional power-law decay (in simplified model calculations: $1/\sqrt{k_F r}$ for 2D superconductors \cite{Zyuzin_2013} and $1/(k_F r)$ for 3D superconductors \cite{pientka_2013}) which significantly decreases the spatial extent of the Majorana end state. We now numerically analyze the spatial extent of the Majorana end modes in our hybrid one-dimensional Fe chain embedded in the superconducting Pb substrate, with
$\Delta$ set to the realistic value 1.3 meV.

In Fig.~\ref{fig:mf_wf} we show one representative wavefunction of the lowest energy eigenstate obtained from exact diagonalizing the Hamiltonian of a finite-size hybrid system. The amplitude of the wavefunction is localized within 20 \AA of the ends of the Fe chain, which is about a couple of Fermi wavelength, and the energy corresponding to this eigenstate is about 0.09 meV, far below the 1.3 meV gap of the host superconductor (also see the highlighted point in Fig.~\ref{fig:sp_comp} upper panel). This is a typical eigenstate in a finite-size system where two Majorana end states are unavoidably coupled \cite{footnote_coupling_energy} -- in this case both through the Fe chain and through the bulk of Pb. The spatial profile of such a state is in good agreement with the experimental observation reported in Ref.~\onlinecite{Nadj-Perge14}, and is reminiscent of the atomic length scale of single Shiba impurity states \cite{Yazdani_1997}. In the lower panel of Fig.~\ref{fig:mf_wf}, we plot the wavefunction on each sublattice of the Fe chain in logarithm scale, which shows clearly a non-exponential decay on top of oscillations associated with the Fermi wavelength. In the inset of the same panel, we further plot the squared amplitude of the wavefunctions scaled by a factor linear in distance \cite{Yazdani_1997}, in order to compare with the predicted $1/\sqrt{r}$ prefactor in the Majorana wavefunction for a 2D host superconductor \cite{Zyuzin_2013}. In fact, we cannot draw a consistent conclusion about the power-law prefactors of the strongly localized (Majorana) eigenstates generically found in the hybird system, possibly because of the involvement of multiple bands in our model and the complicated interplay between the iron states and the Shiba states. Alternatively, another possible explanation for the short localization length of the Majorana end states is the strongly renormalized velocity of the low-energy quasi-particles in the hybrid system \cite{Peng_2014}. A full understanding of these states will be a subject of future work.


The difference between the hybrid system and a purely 1D system is obvious in terms of both their spectra and the wavefunctions. As a reference of a purely 1D system we choose a suspended triple Fe chain with artificially added Rashba spin-orbit coupling (Eq.~\eqref{eq:rashbasimple}) and spin-singlet pairing potential (Eq.~\eqref{eq:hpair}) terms. With exactly the same length of chain and the same realistic $\Delta$, the two systems exhibit a sharp contrast at low energy $E\lesssim\Delta$ (see Fig.~\ref{fig:sp_comp}): in the hybrid system, abundant subgap states can be found from exactly diagonalizing a finite-size Hamiltonian, and the low-energy states generically show localized profiles at the chain ends as illustrated in Fig.~\ref{fig:mf_wf}; in the suspended chain, subgap states are barely seen in a 210-\AA-long chain with a reasonable Rashba spin-orbit coupling strength because of strong finite-size effect -- low-energy states showing pronounced decay from the ends can only occur in such a chain when $\Delta$ is larger than 0.01 eV. This sharp contrast emphasizes the fact that in order to understand various features (including the zero-bias peaks and the short length profile) inside the host superconducting gap reported in Ref.~\onlinecite{Nadj-Perge14}, it is necessary to go beyond a simple 1D (or quasi-1D) model, and to take into account the hybrid nature of the experimental structures. 

\begin{figure}
  \centering
  \includegraphics[width=0.42\textwidth]{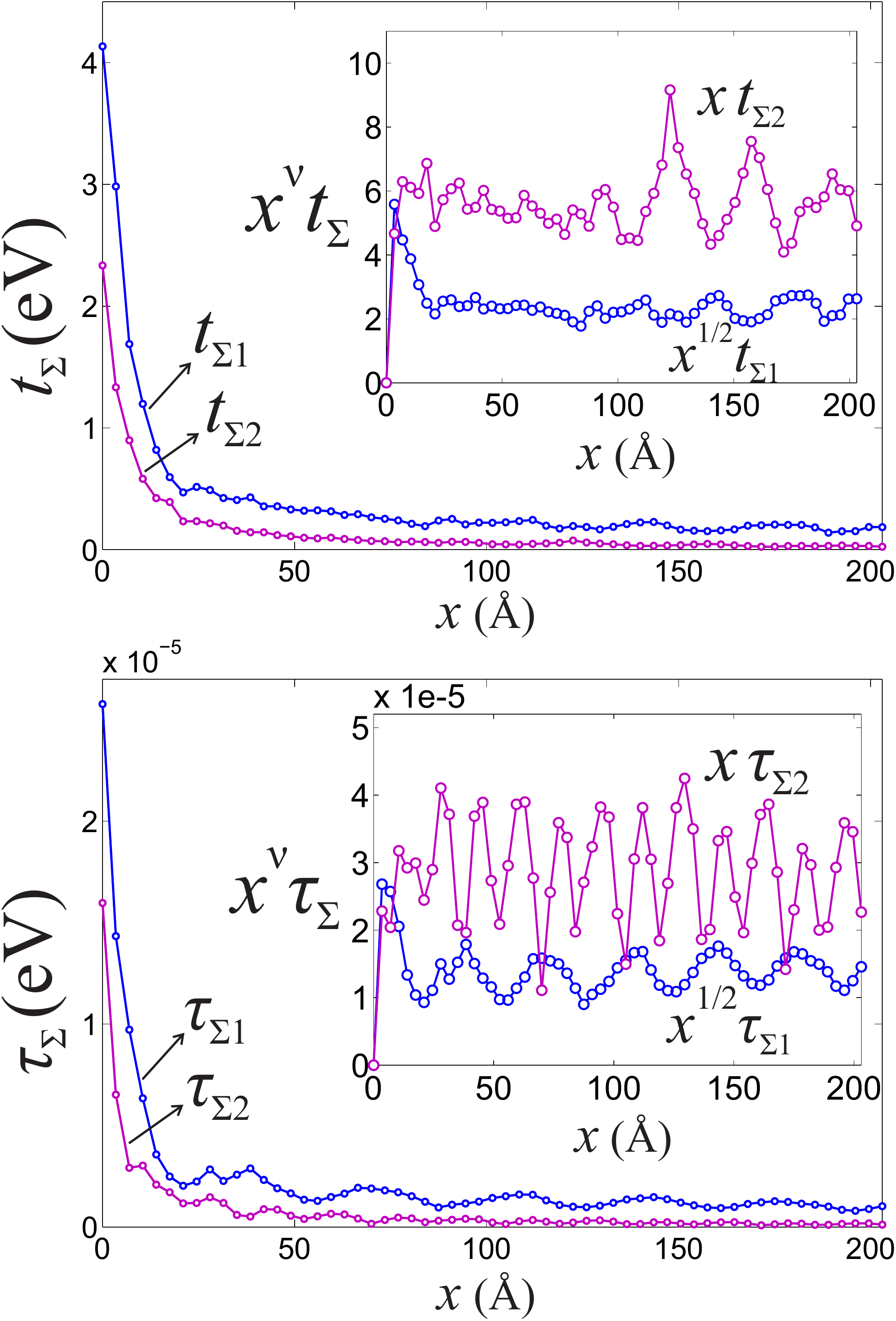}
  \caption{Effective hopping parameters $t_\Sigma$ (upper panel) and inverse-life-time parameters $\tau_\Sigma$ (lower panel; see text for the definitions of these parameters), extracted from the self-energy due to the superconducting substrate (assuming zero energy and $V_{pd\pi} = 1$ eV), as a function of distance $x$. Only two of the largest parameters are shown in each plot. The insets show the same curves scaled by appropriate power-law prefactors in terms of $x$.}
  \label{fig:tse}
\end{figure}

To further reveal the origin of this difference, we return to Eq.~\eqref{eq:Gr_chain} and investigate the properties of the self-energy due to coupling to the superconducting substrate, which effectively differentiates the two systems. To this end we consider a Pb substrate that is infinite in the $xy$ plane and semi-infinite in the $z$ direction. This distinguishes the following results from those in the previous part by removing the finite size effect and recovering the full translational invariance in the self-energy. The self-energy for the Fe chain \cite{footnote_se} is given by [cf. Eq.~\eqref{eq:SE0}]
\begin{align}
  &\Sigma_S(E^+;\bm{r}_1-\bm{r}_2) = \nonumber \\
  &\;\sum_{\bm{r}'_1,\bm{r}'_2} \tau_{\text{Fe-Pb}}(\bm{r}_1-\bm{r}'_1) G_{\text{Pb}}^{(0)}(E^+;\bm{r}'_1-\bm{r}'_2) \tau^\dag_{\text{Fe-Pb}}(\bm{r}_2-\bm{r}'_2),
\end{align}
where $G_{\text{Pb}}^{(0)}$ is the Green function for the Pb substrate in the presence of superconducting pairing and without coupling to the Fe chain. We are particularly interested in the energy range inside the superconducting gap, where $G_{\text{Pb}}^{(0)}$ has no poles. For concreteness we focus on zero energy, and define effective hopping parameters $t_{\Sigma}(x)$ as the singular values of the matrix $[\Sigma_S(x)+\Sigma_S(x)^\dag]/2$, as well as effective inverse-life-time parameters $\tau_{\Sigma}(x)$ as the singular values of $i[\Sigma_S(x)-\Sigma_S(x)^\dag]/2$.

In Fig.~\ref{fig:tse} we plot two of the largest $t_{\Sigma}$'s and $\tau_{\Sigma}$'s as a function of $x$, assuming $V_{pd\pi} = 1$ eV (note that $\Sigma_S\propto V_{pd\pi}^2$). We find that both $t_{\Sigma}(x)$ and $\tau_{\Sigma}(x)$ show long-range characters that can be fitted by either an $x^{-1/2}$ prefactor or an $x^{-1}$ prefactor (see Fig.~\ref{fig:tse} insets). This is reasonable because in the present case we have considered a Pb substrate that is semi-infinite in the $z$ direction; both 2D-like surface states and 3D-like bulk states exist in such a substrate; the effective hoppings that decay as $x^{-1/2}$ have to do with virtual processes through the surface states of the substrate, and those that decay as $x^{-1}$ has to do with virtual processes through the bulk states \cite{pientka_2013}. Moreover, we see that the magnitude of $t_{\Sigma}(x)$ is typically of order 1 eV -- the same as the original hopping parameters in the Fe chain -- when $x$ is within several lattice constants (3.5 \AA). This implies that as long as the coupling between Fe and Pb atoms is sufficiently strong ($V_{pd\pi} > 0.1$ eV), the effective Hamiltonian of the Fe chain will be significantly modified by the self-energy contribution. In particular, if the self-energy contribution becomes dominant, the chain is essentially governed by the physics of the long-ranged Shiba lattice \cite{pientka_2013}. Incidentally, we find that the magnitude of $\tau_{\Sigma}$, corresponding to the intrinsic line-width (or inverse life-time) of the subgap states, is much smaller than that of $t_{\Sigma}$, which is expected as we have focused on the energy far below the superconducting gap.

\begin{figure}
  \centering
  \includegraphics[width=0.47\textwidth]{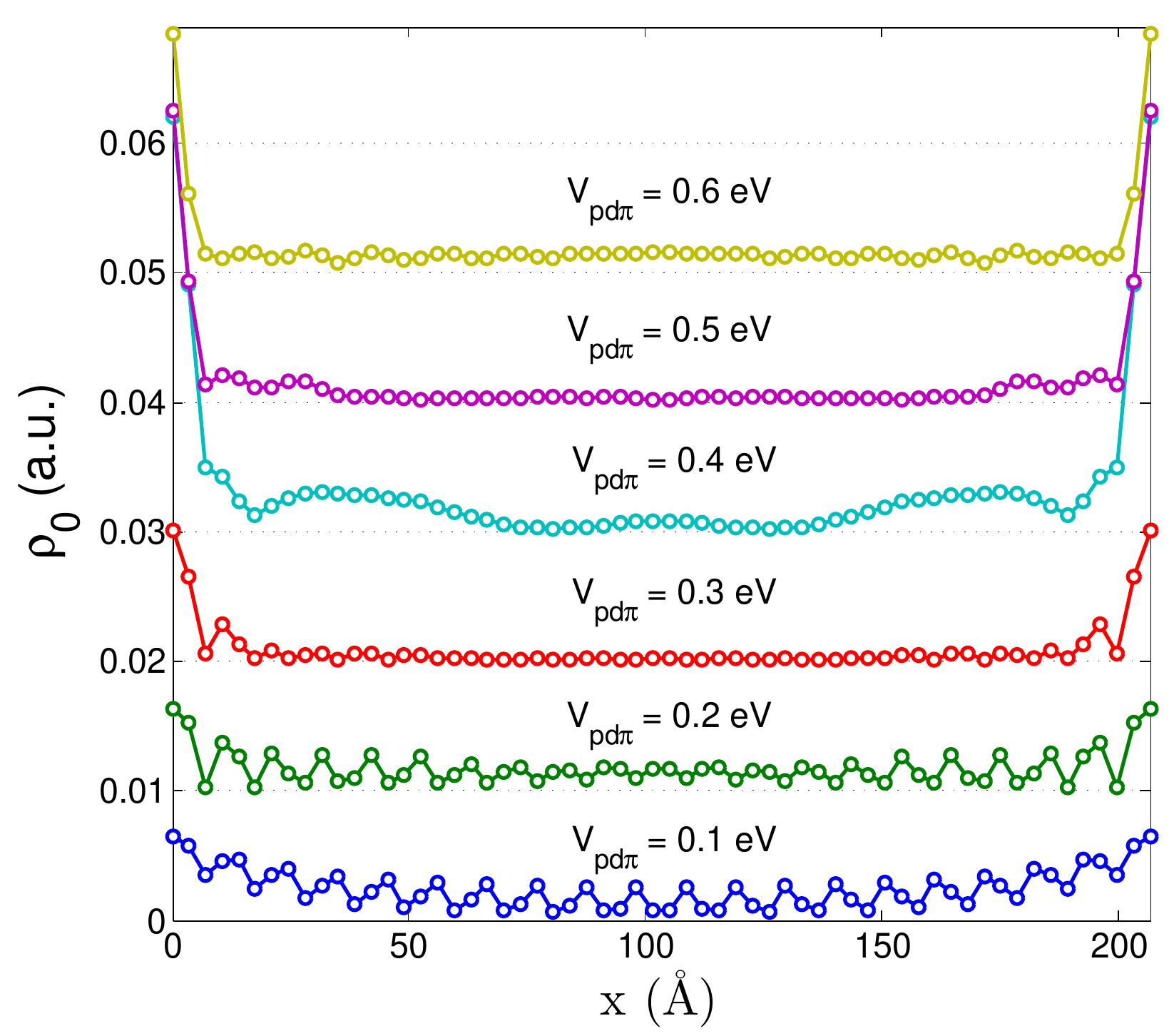}
  \caption{Spatial dependence of the zero-energy spectral functions in a 60-unit-cell-long triple Fe chain, with different values of coupling $V_{pd\pi}$ between Fe and Pb atoms. Data for different $V_{pd\pi}$ are shifted by an increment of 0.01 for clarity. The Pb substrate is infinite in the $xy$ plane and semi-infinite in the $z$ direction, and has a spin-singlet pairing gap $\Delta = 1.3$ meV.}
  \label{fig:spf_Fe}
\end{figure}

As a consequence of the substrate self-energy, not only $p$-wave pairing gaps can be induced in the Fe chain, the spatial profile of Majorana end states are also strongly modified. This can be seen, within the current formalism, in the spatial profiles of spectral functions at zero energy $\rho_0(x) = -\frac{1}{\pi}\mathrm{Im}[\text{Tr}G^r_{Fe}(E=0;x,x)]$, plotted in Fig.~\ref{fig:spf_Fe}. Clearly, when the coupling between the Fe chain and the Pb substrate, characterized by the parameter $V_{pd\pi}$, is strong enough, $\rho_0$ is sharply localized at the ends within a few Fermi wave-length. The spatial profile of such a spectral function does not obey simple exponential decay, and agrees very well with that of the low-energy eigenstates obtained in the finite-size system exemplified previously in Fig.~\ref{fig:mf_wf}, as well as the experimental results in Ref.~\onlinecite{Nadj-Perge14}.

\section{Discussion and Conclusions}

Motivated by recent STM experiments\cite{Nadj-Perge14} which identified zero-bias peaks in the tunneling density-of-states of 
iron atomic chains placed on the surface of lead and interpreted them as Majorana states, we have carried out a 
theoretical study aimed at shedding light on strategies for developing magnetic transition metal atom chains on the surface of superconductors 
as a platform for one-dimensional topological superconductivity. Our conclusions are generally speaking optimistic.
Even though the exchange spin-splitting on the chain typically exceeds the superconductor's Clogston limit 
by orders of magnitude, nano structures of this type typically form gapped superconducting states through a mechanism 
illustrated schematically in Fig.~\ref{fig:schematic}, and these states are often topological.   

Our theoretical study aims to identify some general trends and is not exhaustive. We have restricted our attention to transition metal atomic chains.  
The case of rare earth chains will differ in some important respects and deserves attention. We have also assumed that 
spin-singlet pairing is dominant on the transition metal chains and treated its strength as a phenomenological model 
parameter. Constructing a realistic theory of pairing on the transition metal chain should be feasible, since the 
pairing mechanism is almost certainly dominated by phonon-mediated attractive electron-electron interactions, but still challenging 
in several respects and beyond the scope of the present work. In addition we 
have based our conclusions on models which do not include $s$ orbitals centered on the transition
metal atoms. This omission seems to be justified by two different considerations, namely that $s$-orbitals are 
both more weakly spin-polarized than $d$ orbitals and more strongly dispersive.  Adding $s$-orbitals to the models we have studied will almost always increase the number of bands which cross the normal state Fermi level by two, and will therefore not alter the topological character of the state.    
{Because it is easier to analyze, we have also focused a good part of our attention on the {\em suspended} chain limit in which we 
do not account explicitly for the degrees of freedom of the superconducting substrate and instead integrate them out, retaining only
the pair-potential and the spin-orbit coupling terms induced by virtual occupation of the substrate orbitals.  We have also examined 
the limit of strongly hybridized chains, which appears to be closer to the circumstance examined experimentally 
in Ref.\onlinecite{Nadj-Perge14}.  In particular accounting for strong hybridization with the Pb substrate explains
the experimental finding that the chain end states are weakly coupled and strongly localized in space, even though 
these studied chains were shorter than estimated superconducting coherence lengths.}

There is a strong interplay between the possibility of achieving topological superconductivity in transition metal atom chains and 
the nature of the magnetic order in these chains. In this paper we have addressed the case of ferromagnetic chains 
with an easy magnetic axis perpendicular to the chain. We have therefore been assuming that the 
magnetic stiffness along the chain is sufficiently large to justify a macrospin limit and that the magnetic anisotropy 
is sufficiently strong that the overall spin-orientation does not suffer thermal fluctuations.
For Fe on Pb these conclusions are supported by {\it ab initio} electronic structure calculations. There is in fact a large experimental and theoretical literature on magnetic order in 
one-dimensional chains.\cite{chain_Gambardella_2002,chain_Hirjibehedin_2006,chain_Serrate_2010, chain_Smogunov_2008, chain_schubert_2011, chain_tung_2007, chain_zeleny_2009}
Magnetism is influenced by bond lengths, bond angles, band fillings, and 
substrate among other factors. Chains made from elements that are magnetic in the bulk do not necessarily have ferromagnetic order and 
conversely, chains made from elements that are not magnetic in the bulk can be ferromagnetic as a chain. It is generally a nontrivial task to experimentally determine the magnetic order of a specific chain.
On the theory side, {\em ab initio} density functional theory calculations can be helpful in identifying the magnetic order of a chain once its structure and composition is known. Generally speaking transition metal atom chains tend to be ferromagnetic when the atoms are close together and antiferromagnetic when the atoms are far apart. The magnetic interactions in these limits can be 
interpreted as being dominated by double exchange and super exchange respectively. Helical and other more complex textures tend to occur
close to the crossover between these limits. For the particular case of transition metal atoms on lead, however, strong 
$p-d$ bonding leads to closely spaced transition metal atoms.  We expect simple ferromagnetism in nearly every case and 
this has motivated our restriction to uniform exchange fields.

Our model studies have allowed us to reach two main conclusions which will, we hope, inform efforts to develop ferromagnetic chains on superconducting substrates as a practical Majorana state factory.  
\begin{itemize}

\item
 i) Pb is an excellent superconducting substrate. It is a relatively large gap superconductor. Its $p$-orbitals readily hybridize with $d$-orbitals in the transition metal chain allowing Cooper pairs to hop from the substrate to the 
 magnetic chain. Its strong spin-orbit coupling not only provides the Rashba spin-orbit coupling required for gapped superconducting states, but also has a favorable influence on chain magnetic properties by enhancing the chain magnetocrystalline anisotropy and by inducing Dzyaloshinskii-Moriya (DM) interactions \cite{dzyaloshinskii_1958,moriya_1960} between the chain magnetic atoms. A large magnetocrystalline anisotropy stabilizes the magnetic order of the chain and is generally desirable. The DM interactions can lead to canted/spiral magnetic order, which, in combination with the on-site spin splitting, can contribute to the effective Rashba spin-orbit coupling.\cite{stevan_2013,klinovaja_2013} 

\item
ii) The iron atoms chains studied in Ref.\onlinecite{Nadj-Perge14} probably do not optimize ferromagnetic chain topological superconductivity. This conclusion motivates a program of experimental and theoretical research aimed at forming topological superconductors with the largest possible gaps and the most robustly 
reproducible topological character. Our model calculations indicate, for example, that the superconducting state is most likely to be topological when the ferromagnetic atom chain is straight. The structure formed by ferromagnetic atoms on lead is influenced both by the mixture of atoms that are present and by the chain growth conditions. If protocols can be established for growing straight chains, they should enable perfectly reproducible topological
behavior. Our model calculations indicate that the one-dimensional superconducting gap is $\sim \Delta E_{so}/J$. $\Delta$ and $E_{so}$ should be enhanced by strong hybridization with a strongly spin-orbit coupled superconductor like Pb. This formula indicates however that larger superconducting gaps might be achievable in chains with itinerant electron ferromagnetism that is weaker than in iron, perhaps in a chain formed by atoms that are not magnetic in the bulk and barely magnetic in the less coordinated chain geometry.  

\end{itemize} 

In summary, we carried out a study of topological superconductivity and Majorana end states in $3d$ ferromagnetic chain tight-binding models with spin-orbit coupling, inversion symmetry breaking, and $s$-wave superconductivity pairing. 
We found that the atomic spin-orbit coupling is in general not sufficient for a $p$-wave superconducting gap to be opened in the ferromagnetic chain, and that one needs to break inversion symmetry or introduce Rashba spin-orbit coupling. 
This property can be explained with an argument similar to that used for the 3D Weyl semimetals. Motivated by recent experiments, we discussed in detail how a sizable Rashba spin-orbit coupling is induced in the ferromagnetic chain when it is deposited on a strongly spin-orbit coupled substrate. We have constructed topological phase diagram in model parameter spaces, varying band filling, exchange splitting strength, and chain structural parameters. In straight magnetic chains we found that the half metallicity which appears at 
strong exchange splitting makes topological superconductivity particularly robust, especially compared to the case of the semiconductor quantum wire Majorana platform. Finally we discussed the possible appearance of a new symmetry protecting an integer number of Majorana modes (where interaction effects could potentially be seen), and highlighted the crucial role that the hybrid structure plays in the decay of the Majorana end modes.  

{\it Note added}. We thank Felix von Oppen and Falko Pientka for helpful discussions and for sharing with us their unpublished results.

\begin{acknowledgments}
H.C. and A.H.M. were supported by the Welch Foundation under Grant No. TBF1473 and by the Office of Naval Research under grant ONR-N00014-14-1-0330. A. Y. acknowledges support from ONR-N00014-14-1-0330, ONR-N00014-11-1-0635, ONR- N00014-13-10661, NSF-MRSEC programs through the Princeton Center for Complex Materials DMR-0819860, NSF-DMR-1104612, ARO-W911NF-1-0262, ARO-MURI program W911NF-12-1-0461, DARPA-SPWAR Meso program N6601-11-1-4110. BAB acknowledges support from  NSF CAREER DMR-0952428, ONR-N00014-11-1-0635, MURI-130-6082, Packard Foundation, and Keck grant. J. L. acknowledges support of Swiss National Science Foundation. H.C. is grateful to Chih-Kang Shih for helpful discussions.
\end{acknowledgments}

\appendix
\section{Calculation of Majorana number}
To calculate the Majorana number (defined below), we need first to write the BdG Hamiltonian into Majorana fermion basis. For any fermion operator $\psi$ one can define two Majorana operators
\begin{eqnarray}
\gamma_a=\psi+\psi^\dag,\,\,\,\gamma_b=-i(\psi-\psi^\dag).
\end{eqnarray}
The Majorana operators fulfill the following relations
\begin{eqnarray}
&&\gamma^\dag_{i,\alpha}=\gamma_{i,\alpha},\,\,\, \alpha=a,b\\\nonumber
&&\{\gamma_{i,\alpha},\gamma_{j,\beta}\}=2\delta_{ij}\delta_{\alpha\beta}.
\end{eqnarray}
By explicitly writing all possible terms in a quadratic fermionic Hamiltonian in the Majorana basis, and considering the hermicity of the coefficients, we can prove that any quadratic fermionic Hamiltonian up to a constant can be written as
\begin{eqnarray}\label{eq:Hmajoranar}
H=\frac{i}{2}\sum_{ij}\gamma_i^T A_{ij} \gamma_j,
\end{eqnarray}
where $\gamma_{i}\equiv (\gamma_{i,a},\gamma_{i,b})^T$, and the matrix $A$ is real and antisymmetric.

The Fourier transform of Majorana fermions is 
\begin{eqnarray}
\gamma_{i,\alpha} = \frac{1}{\sqrt{N}} \sum_{i} e^{-i\bm k\cdot \bm r_i} \gamma_{\bm k,\alpha},
\end{eqnarray}
which implies that $\gamma^\dag_{\bm k,\alpha}=\gamma_{-\bm k,\alpha}$. The Hamiltonian after Fourier transform becomes
\begin{eqnarray}\label{eq:Hmajoranak}
H=\frac{i}{2}\sum_{\bm k}\gamma_{\bm k}^\dag \tilde{A}_{\bm k} \gamma_{\bm k},
\end{eqnarray}
and $\tilde{A}_{\bm k}$ is still antisymmetric but not necessarily real. Eq.~\ref{eq:Hmajoranak} is closely related to the BdG Hamiltonian written in Nambu spinors. Below we follow the prescription given in Ref.~\onlinecite{budich_2013}. 

In real space the Majorana spinor $\gamma_i = (\gamma_{i,a},\gamma_{i,b})^T$ is related to the Nambu spinor $\Psi_i = (\psi_{i},\psi^\dag_{i})^T$ by 
\begin{eqnarray}
\gamma_i = \sqrt{2} U \Psi_i,\,\,\, U= \frac{1}{\sqrt{2}} \left ( \begin{array}{cc}
1 & 1\\
-i & i\\
\end{array} \right )
\end{eqnarray}
Therefore after Fourier transform we have
\begin{eqnarray}
\tilde{A}_{\bm k} = -iU H_{\rm BdG}(\bm k) U^\dag,
\end{eqnarray}
which makes it convenient to obtain $\tilde{A}_{\bm k}$ from the BdG Hamiltonian.  

The Majorana number $\mathcal{M}$ of an infinite 1D chain is defined as
\begin{eqnarray}
\mathcal{M} = {\rm sgn}\left [  {\rm Pf} ( \tilde{A}_{\bm k=0}  ) {\rm Pf} ( \tilde{A}_{\bm k=\frac{\pi}{a}} )    \right ],
\end{eqnarray} 
where $a$ is the lattice constant, and $\rm Pf$ means Pfaffian of an antisymmetric matrix. When $\mathcal{M}=-1$, i.e., ${\rm Pf} ( \tilde{A}_{\bm k}  )$ takes opposite signs at zone center and zone boundary, the chain is topologically nontrivial and there should be isolated zero energy Majorana edge modes in a finite chain.


\begin{thebibliography}{99}
\bibitem{footnote_anisotropy} This magnetization direction is favored by magnetostatic interactions 
and for the Fe case, also by spin-orbit coupling.  See Ref.\onlinecite{Nadj-Perge14}.
\bibitem{footnote_coupling_energy} Note that the coupling energy here is still smaller than the highest energy resolution in the experiments of Ref.~\onlinecite{Nadj-Perge14}, which is about 0.1 meV.
\end{thebibliography}
\end{document}